\documentclass{emulateapj}
\newcommand{\bm}[1]{\text{\boldmath$\mathit #1 $}}

\usepackage{amsmath,bm}

\shorttitle{SN Ia NIR Hierarchical Bayesian Inference}
\shortauthors{Mandel et al.} 

\begin{document}
\title{Type Ia Supernova Light Curve Inference: \\Hierarchical  Bayesian Analysis in the Near Infrared}
\author{Kaisey S. Mandel\altaffilmark{1}, W. Michael Wood-Vasey\altaffilmark{2}, Andrew S. Friedman, Robert P. Kirshner}
\affil{Harvard-Smithsonian Center for Astrophysics, 60 Garden St., Cambridge, MA 02138}
\altaffiltext{1}{kmandel@cfa.harvard.edu}
\altaffiltext{2}{Current Address: Dept. of Physics \& Astronomy, 100 Allen Hall, 3941 O'Hara St., Univ. of Pittsburgh, Pittsburgh, PA 
15260}

\slugcomment{Submitted 30 Nov 2008;  ApJ Accepted 31 Jul 2009}

\begin{abstract}
We present a comprehensive statistical analysis of the properties of Type Ia SN light curves in the near infrared using recent data from 
PAIRITEL and the literature.   We construct a hierarchical Bayesian framework, incorporating several uncertainties including 
photometric error, peculiar velocities, dust extinction and intrinsic variations, for principled and coherent statistical inference.  SN Ia 
light curve inferences are drawn from the global posterior probability of parameters describing both individual supernovae and the 
population conditioned on the entire SN Ia NIR dataset.   The logical structure of the hierarchical model is represented by a 
directed acyclic graph.  Fully Bayesian analysis of the model and data is enabled by  an efficient  MCMC algorithm exploiting the 
conditional probabilistic structure using Gibbs sampling.  We apply this framework to the $JHK_s$ SN Ia light curve data.  A new light 
curve model captures the observed $J$-band light curve shape variations.   The marginal intrinsic variances in peak absolute 
magnitudes are: $\sigma(M_J) = 0.17 \pm 0.03$, $\sigma(M_H) = 0.11 \pm 0.03$, and $\sigma(M_{Ks}) = 0.19 \pm 0.04$.  We describe 
the first quantitative evidence for correlations between the NIR absolute magnitudes and $J$-band light curve shapes, and 
demonstrate their utility for distance estimation.  The average residual in the Hubble diagram for the training set SN at $cz > 2000 
\text{ km s}^{-1}$ is 0.10 mag.  The new application of bootstrap cross-validation to SN Ia light curve inference tests the sensitivity of the 
statistical model fit to the finite sample and estimates the prediction error at 0.15 mag.  These results demonstrate that SN Ia NIR light 
curves are as effective as corrected optical light curves, and, because they are less vulnerable to dust absorption, they have great potential as precise and accurate  cosmological distance indicators. 

\end{abstract}


\section{Introduction}

Type Ia supernova (SN Ia) rest-frame optical light curves have been of great utility for measuring of fundamental quantities of the 
universe.  As standardizable candles, they were critical to the detection of  cosmic acceleration \citep{riess98,perlmutter99}.   The 
cosmic acceleration may be caused by a dark energy component of the universe (\citealt{frieman08} provide a recent review).  SN Ia  
have been used to constrain the equation-of-state parameter $w$ of dark energy  \citep{garnavich98b}, and recent efforts have 
measured $w$ to 10\%, \citep{wood-vasey07, astier06, kowalski08, hicken09b}.   SN Ia have also been used to establish the 
extragalactic distance scale and measure the Hubble constant \citep{freedman01,jha99,riess05,riess09a,riess09b}, 

The effectiveness of SN Ia as distance indicators has been improved greatly by the construction of empirical methods that exploit 
relationships between peak optical luminosities of SN Ia and distance-independent measures such as light curve shape or color that 
have been observed in the burgeoning sample of nearby low-$z$ SN Ia \citep{hamuy96_29sne, riess99, jha06, hicken09a}.  Methods 
have included $\Delta m_{15}(B)$ \citep{phillips93,hamuy96, phillips99}, MLCS \citep{riess96, riess98, jha07}, ``stretch'' 
\citep{goldhaber01}, CMAGIC \citep{wang03}, and SALT \citep{guy05, guy07}.  The largest systematic uncertainty that limits the 
precision of rest-frame optical light curves is dust extinction in the host galaxy and the entanglement of dust reddening with the intrinsic 
color variations of SN \citep[e.g.,][]{conley07}.

Early observations of SN Ia in the infrared were made by \citet{kirshner73, elias81, elias85, frogel87} and \citet{graham88}.  Studies of nearby SN Ia 
light curves in the NIR  have found the peak near-infrared luminosities of SN Ia have a dispersion smaller than $\pm 0.20$ mag  
\citep{elias85, meikle00, krisciunas04a,krisciunas04c}.  Furthermore, the effect of dust extinction is significantly diminished at near-infrared wavelengths, relative to the optical.  The combination of optical and near-infrared observations of SN Ia light curves 
could lead to even better SN Ia distances \citep{krisciunas07}.

\citet{wood-vasey08} (hereafter WV08) compiled the largest homogeneous sample of NIR SN Ia observations, taken with the Peters 
Automated InfraRed Imaging TELescope \citep[PAIRITEL;][]{bloom06b}.  After combining these with NIR light curve observations from 
the literature to yield a sample of 41 NIR SN Ia, they constructed template light curves by interpolating and smoothing the data.
They measured the scatter in the absolute magnitudes at time of $B$ maximum in each of the $J$, $H$, and $K_s$ bands and found $
\sigma(M_H) \approx 0.15$ mag, $\sigma(M_J) \approx 0.33$ mag and $\sigma(M_{K_s}) \approx 0.26$ mag.   This analysis did not 
take into account NIR light curve shape variations, but it was found, as in \citet{krisciunas04a}, that the Hubble diagram residuals had 
no trend with the optical light curve width.

The purpose of this paper is twofold.  First, we formulate the hierarchical Bayesian approach to probabilistic inference with SN Ia light 
curves in general.  A proper Bayesian approach provides a principled, coherent framework for inference based on the  joint probability 
density over all quantities of interest conditioned on the available data.   It is natural to employ a simultaneous multi-level approach 
and derive joint probability densities over the parameters of individual supernova light curves, their distance moduli, and also the 
variables that describe the population, including those governing the joint probability distributions over multi-band absolute 
magnitudes and light curve shape parameters.   This approach enables consistent statistical inference of all hierarchical parameters, 
by coherently incorporating several sources of uncertainty, including peculiar velocity uncertainties, photometric measurement errors, 
intrinsic randomness of light curves, and dust extinction into the global posterior probability density conditioned on the entire data set 
simultaneously.  This framework leads to a natural and consistent method for probabilistic distance prediction with new SN Ia light 
curve data.   The logical structure of our hierarchical model for SN Ia light curve inference is demonstrated by the equivalent directed acyclic 
graph (DAG), a graphical model that facilitates easy inspection of the probabilistic relationships.

Although the probabilities for fully Bayesian analysis of SN Ia light curves are simple to write down, the joint posterior distribution is 
generally non-gaussian and  difficult to evaluate.  To enable probabilistic inference, we have developed a Markov Chain Monte Carlo 
(MCMC) algorithm, \textsc{BayeSN}, designed to exploit the conditional probabilistic structure using Gibbs sampling.  We employ this code for 
both  training the statistical model and  using the model to predict distances.  The use of advanced sampling methods facilitates the 
computation of marginal probabilities of parameters from the global joint posterior density.

In the second part of the paper, we apply this framework to the NIR  SN Ia light curve data from the compilation of WV08.  We first 
construct model template light curves for the $JHK_s$ bands.  We compute fixed maximum likelihood template models between -10 
and 20 days for the $H$ and $K_s$ using all available data.  The $J$-band data is typically much less noisy than the $H$ and $K$, so 
we construct an extensive $J$-band model between -10 and 60 days that accounts for light curve variations, in particular the 
structure around the second maximum.  Next, we apply the \textsc{BayeSN} method to simultaneously (1) fit the individual $JHK_s$ 
light curves, (2) compute the population characteristics, especially the absolute magnitude variances and covariances with $J$-band 
light curve shape and (3) estimate the joint and marginal uncertainties over all hierarchical parameters.  We construct a Hubble 
diagram for the training set SN Ia and compute its residual errors.  

The average Hubble diagram residual of the training set SN is an optimistic assessment of the predictive ability of the statistical model 
for SN Ia light curves because it uses the SN data twice: first for estimating the model parameters (training), and second in evaluating 
the error of its ``predictions''.  Hence, the residuals, or training errors, underestimate the expected prediction error.  This effect is 
present for all models based on finite training data, and is particularly important for small sample sizes.   We perform bootstrap cross-
validation to realistically estimate the out-of-sample prediction error and to test the sensitivity to the finite NIR SN Ia sample.   This 
technique ensures that the same SN are not simultaneously used for  training and prediction.  It has not been used previously in SN Ia 
statistical modeling and inference.

This paper demonstrates hierarchical Bayesian modeling and distance estimation for SN Ia light curves in the NIR only.  The application of these methods to combined optical and NIR light curves for the estimation of dust and distances will be described in a subsequent paper (Mandel et al. 2009, in prep.).

This paper is organized as follows:  In \S 2, we describe the hierarchical Bayesian framework for SN Ia light curve inference.  The 
structure of the hierarchical model can be depicted formally as a directed acyclic graph presented in \S 2.3.   In \S 2.4, we describe \textsc{BayeSN}, 
an MCMC algorithm designed for computing posterior inferences in the hierarchical framework.   In \S 3, the construction of 
template light curve models in $JHK_s$ is described.  In \S 4, we apply this approach to the NIR light curve data, and summarize the 
posterior inferences for both individual supernovae and the population.   In \S 4.4, we construct Hubble diagrams by applying the 
statistical model and describe the application of bootstrap cross-validation to estimate prediction error.  In \S 4.5, we discuss the 
potential impact of dust in the NIR sample.   We conclude in \S 5.  In appendix \S \ref{dsep}, we briefly review the conditional independence properties of graphical models.  Appendix \S \ref{bayesnmath} presents mathematical details of the \textsc{BayeSN} method, and appendix \S \ref{practical} describes its use in practice.

\section{Hierarchical Bayesian Framework for SN Ia Light Curve Inference}

Simple Bayesian analysis describes an inference problem in which a generative model $\mathcal{H}$ with a free parameter $\theta$ 
is assumed to underly the observed data $\mathcal{D}$.  The Bayesian paradigm is to derive inferences on $\theta$ from the posterior 
density of the parameter conditioned on the data: $
P(\theta | \, \mathcal{D}, \mathcal{H}) \propto P( \mathcal{D} | \, \theta, \mathcal{H}) P( \theta | \mathcal{H})$,
where the first factor  is the likelihood function and the second factor is the prior on the model parameter. 

Hierarchical, or multi-level, Bayesian analysis is a modern paradigm of statistical modeling, which enables the expression of rich 
probabilistic models with complex structure on multiple logical levels \citep{gelman_bda}.  For example, if $\mathcal{D}_i$ represents 
the data on \emph{individual} $i$, and $\theta_i$ is a model parameter describing $i$, the values of $\theta_i$ themselves may be 
drawn from a prior or \emph{population} distribution $P( \theta_i | \, \alpha, \beta)$, which in turn depends on unknown variables that 
describe the \emph{group} level probabilistic model.  These unknown variables $\alpha, \beta$ are termed \emph{hyperparameters} to 
distinguish them from the individual level parameters $\theta_i$.  The hierarchical Bayesian joint posterior distribution over all 
parameters $\{ \theta_i \}$ and hyperparameters $\alpha, \beta$ conditioned on the data for many ($N$) individuals $\mathcal{D} = \{ \mathcal{D}_i \}$ is then:
\begin{equation}
P(\{\theta_i\}; \alpha, \beta | \, \mathcal{D}) \propto \left[ \prod_{i=1}^N P(\mathcal{D}_i | \, \theta_i) P(\theta_i | \alpha, \beta) \right] 
P(\alpha, \beta)
\end{equation}
where the last factor represents the \emph{hyperprior} density on the hyperparameters.  The fully Bayesian approach is to analyze the 
full joint posterior density of all parameters and hyperparameters simultaneously conditioned on the entire data set.  This ensures the 
complete and consistent accounting of uncertainty over all the inferred parameters.
We can build in complex probabilistic structure by layering single-level models, at the individual level and also at possibly multiple 
population levels, and expressing the conditional relationships that connect them.  The hierarchical Bayesian paradigm is very well 
suited to combining information and uncertainties from many logical sources of randomness, interacting in non-trivial ways, in a 
principled, coherent and consistent statistical framework for studying structured data.  This is the strategy we adopt in this paper.

In contrast to more classical methods of model fitting, the Bayesian approach is less concerned with the optimization problem, i.e. 
finding the ``best'' values or point estimates of model parameters fitted to given data, and more concerned with the construction of the 
full joint probability model,  consistent integration over uncertainties, and the simulation and sampling of the joint posterior density.  For 
most non-trivial models, the integrations of interest are  analytically intractable, so we employ modern computation techniques to 
enable probabilistic inference.   We introduce an MCMC algorithm (\textsc{BayeSN}) that uses stochastic simulation to 
calculate posterior inferences from the hierarchical framework.

\subsection{SN Ia Light Curve Models}

A central task in the statistical analysis of Type Ia SN light curve data is fitting empirical light curve models to time-series 
photometric data in multiple passbands.  
A data set for an individual supernova $s$, $ \{\mathcal{D}^F \} $, consists of observations in $n$ photometric filters, $F \in \{ F^1, \ldots, 
F^n\}$ (e.g. $\{B,V,J,H\}$).  Let $\mathcal{D}^F = \{ t_i, m^F_i, \sigma_{F,i}^2 \}_{i=1}^{N_F}$ be the set of $N_F$ observations in band 
$F$.   We assume these have already been corrected for time dilation \citep{blondin08} and $K$-corrected to the supernova rest 
frame, so that $t_i$ is the rest-frame phase of the observation referenced to the time of maximum light in $B$-band (i.e. $t=0$ 
corresponds to $B_\text{max}$), $m_i^F$ is the apparent magnitude in filter $F$ at this phase, and $\sigma^2_{F,i}$ is the 
measurement variance.  

We wish to fit to this data  a light curve model, $F_0 + l^F(t, \bm{\theta}^F)$, where $F_0$ is the $F$-band apparent magnitude of the 
model at a reference time $t=0$, and $l^F(t,\bm{\theta}^F)$ is the \emph{normalized} light curve model such that $l^F(0, \bm{\theta}^F) = 0$.    A 
vector of \emph{light curve shape parameters}, $\bm{\theta}^F$, governs the functional form of the light curve models and may take on 
different values for each supernova.

At the present time there is no known set of physical functions $l^F(t, \bm{\theta}^F)$ describing the temporal evolution of supernovae 
light curves.  Theoretical SN Ia models alone provide insufficient guidance, so these functions must be constructed from the 
light curve data itself.  Typical parameterizations are motivated by a combination of simplicity, intuition, mathematical convenience and 
the empirical study of detailed light curve data of a large sample of supernovae.  The ultimate utility of  a 
functional form lies in its ability to fit the observed modes of variation in data and capture the observable information in the light 
curves.  Examples of light curve shape functions are:
\begin{enumerate}
\item ``Stretch'' template method \citep{perlmutter99,goldhaber01}:  $l^F(t, \theta^F) =  f( s^F t)$, where the shape parameter is $
\theta^F = s^F$, the ``stretch'' factor in filter $F$, and $f(\cdot)$ is a fiducial light curve, e.g. the Leibundgut template 
\citep{leibundgut89}, which is a function of the rest-frame supernova phase with respect to $t_0$, the time of $B_{\max}$.
\
\item The $\Delta m_{15}(B)$ decline rate parameterization \citep{hamuy96, phillips99}.  Here the light curve shape parameter is $
\theta^B = \Delta m_{15}(B)$, the magnitude decline between the peak and 15 days after the peak in $B$-band.  The light curve function is defined 
at particular values of $\theta^B_i$ using $BVI$ light curve templates generated from observed supernovae.   Interpolation is used to fit light curves at intermediate values of $\Delta m_{15}(B)$.     \citet{prieto06} presented an updated 
formulation.

\item MLCS \citep{riess96, riess98, jha07}:  The light curve model in $UBRVI$ is of the form: $l^F(t, \Delta) = F_0(t) + \Delta P_F(t) + 
\Delta^2 Q_F(t)$, where $F_0(t)$, $P_F(t)$, and $Q_F(t)$ are defined by templates.   The light curve shape parameter is $\theta^F = 
\Delta$.
\end{enumerate}

In this section, we do not assume any particular form for the light curve models $l^F(t; \bm{\theta}^F)$.   The results will be applicable 
to a broad class of possible models.    Without loss of generality, light curve model can depend on some parameters ($
\bm{\theta}^F_{\text{L}}$) linearly and others ($\bm{\theta}^F_{\text{NL}}$) nonlinearly.  Hence, a general form of a light curve model in 
band $F$ for the data is
\begin{equation}\label{gen_lc}
m^F_i = F_0 + l^F_0(t _i; \bm{\theta}^F_{\text{NL}}) + \bm{l}^F_1(t_i ; \bm{\theta}^F_{\text{NL}}) \cdot \bm{\theta}^F_{\text{L}} + 
\epsilon^F_i .
\end{equation}
where $\bm{l}^F_1(t; \bm{\theta}^F_{\text{NL}})$ is a vector of coefficients to the linear parameters.  It is convenient for computational 
purposes to separate the linear from nonlinear shape parameters.  The parameter $F_0$ could be considered a linear parameter and 
included with $\bm{\theta}_{\text{L}}$.  However, they are physically  distinct quantities, as $F_0$ sets the apparent magnitude scale for the 
light curve whereas $\bm{\theta}^F_{\text{L}}$ generally models the shape of the light curve, so we keep them separate.  

\subsection{Constructing the Global Posterior Density}\label{constructposterior}

\subsubsection{Light Curve Likelihood Function}

Assuming  Gaussian noise for $\epsilon_i^F$, we can write the likelihood of the light curve model parameters $ 
\bm{\theta}^F = (\bm{\theta}^F_{\text{L}}, \bm{\theta}^F_{\text{NL}})$ for a single SN data set $\mathcal{D}^F$ in one band $F$.  Define the vectors and 
matrices:
\begin{equation}
\bm{m}^F = (m^F(t_1), \ldots, m^F(t_{N_F}) )^T
\end{equation}
\begin{equation}
\bm{L}^F_0(\bm{\theta}^F_{\text{NL}}) = (l_0^F(t_1; \bm{\theta}^F_{\text{NL}}),\ldots, l_0^F(t_{N_F}; \bm{\theta}^F_{\text{NL}}) )^T
\end{equation}
\begin{equation}
\bm{L}_1^F(\bm{\theta}^F_{\text{NL}}) = (\bm{l}^F_1(t_1; \bm{\theta}^F_{\text{NL}}),\ldots, \bm{l}^F_1(t_{N_F}; 
\bm{\theta}^F_{\text{NL}}))^T.
\end{equation}
Let us construct a vector of ones, $\bm{1}$, of the same length as the data $\bm{m}_s^F$, and a measurement error covariance matrix 
$\bm{W}^F$.  Due to the practical difficulties of estimating the error covariances,  the current standard assumption is that the error 
terms $\epsilon_i^F$ are independent, so that $W_{ii}^F = \sigma^2_{F,i}$ is diagonal.  The likelihood function for the light curve can 
be compactly written as:
\begin{equation}
\begin{split}
 P&( \mathcal{D}^F | \, F_0, \bm{\theta}^F ) = \\
 &N(  \bm{m}^F | \, \bm{1}F_{0} + \bm{L}_0^F(\bm{\theta}^F_{\text{NL}}) + \bm{L}_1^F(\bm{\theta}^F_{\text{NL}}) \cdot 
\bm{\theta}_{\text{L}}^F ,  \bm{W}^{F})
\end{split}
\end{equation}
where $N(\bm{x} | \, \bm{\mu}_x, \bm{\Sigma}_x)$ denotes the multivariate normal density in the random vector $\bm{x}$ with mean $
\bm{\mu}_x$ and covariance $\bm{\Sigma}_x$.
Since the photometric observations in multiple filters are sampled with independent noise, the likelihood function of all light curve 
parameters over all bands given the multi-band data $\{\mathcal{D}^F\}$ is the simple product of $n$ single-filter likelihoods.
\begin{equation}\label{likelihood}
P(\{\mathcal{D}^F \}| \, \bm{\phi}) =  \prod_F P(\mathcal{D}^F | \, F_{0} , \bm{\theta}^F )
\end{equation}
We define the \emph{observable} (or apparent) parameters vector 
$\bm{\phi} = ( F_0^1 , \ldots , F_0^n ; \, \bm{\theta}^{F^1}, \ldots,  \bm{\theta}^{F^n})$.
This vector, with the light curve model functions $l^F(t; \bm{\theta}^F)$, encodes all the information needed to reconstruct the 
\emph{apparent} light curve of a single supernova, i.e. the apparent magnitudes at the reference time and the light curve shape 
parameters in each of $n$ photometric bands $F$.
Similarly, we define the  \emph{intrinsic} (or absolute) parameters vector $\bm{\psi} = (M_{F^1}, \ldots, M_{F^n};\, \bm{\theta}^{F^1},
\ldots, \bm{\theta}^{F^n})$ encoding all the information describing the \emph{absolute} light curves of the supernova.  The absolute 
magnitude at peak in filter $F$ is $M_F = F_0 - \mu - A_F$, where $\mu$ is the distance modulus and $A_F$ is the dust absorption in 
that filter, for a particular supernova.

\subsubsection{Redshift-Distance Likelihood Function}

Type Ia SN can be used as distance indicators because we possess some knowledge of their \emph{relative} distances in the low 
redshift regime (where they are independent of the cosmological parameters $\Omega_M, \Omega_\Lambda$, and $w$) from the 
Hubble law and measured redshifts to the host galaxies of the SN.  However, inference of true luminosities and absolute distances 
requires external calibration (e.g. from Cepheids) or knowledge of the Hubble constant, $H_o$, which has not yet been independently 
measured to high precision.  If we are concerned only with relative distance estimation, it is sufficient to fix the distance scale with an 
assumed $h = H_o / 100 \text{ km s}^{-1}$.  The uncertainty in these local distances is then dominated by the peculiar velocity field, 
which we model probabilistically.  

Let $z_c$ be the cosmological redshift of a SN.  The measured redshift is $z$, with measurement variance $\sigma^2_z$, corrected to 
the CMB and the local infall flow model of \citet{mould00}.    In a smooth cosmological model, the distance modulus is related to $z_c$: 
$\mu = f(z_c) = 25 + 5 \log_{10}[d_\text{L}(z_c) \, \text{Mpc}^{-1}]$, where $d_\text{L}(z_c)$ is the luminosity distance in Mpc.  If we 
model the effect of random peculiar velocity as a Gaussian noise with variance $\sigma_{\text{pec}}^2$, then $z = z_c + N(0, 
\sigma^2_{\text{pec}}/c^2 + \sigma^2_z)$, and the likelihood function is $P(z |\, \mu) = N(z | \, f^{-1}(\mu), \sigma^2_{\text{pec}}/c^2 + 
\sigma^2_z)$.   The posterior density of the distance modulus conditioning only on the redshift is $P(\mu | z) \propto P(z | \mu)$. We 
use a flat prior $P(\mu) \propto 1$, since we have no \emph{a priori} knowledge about $\mu$ without the data.
For recession velocities $cz \gg \sigma_{\text{pec}}$, $f(z_c)$ can be linearized about the fixed $z$ to find $f^{-1}(\mu)$, so that, to a 
good approximation,
\begin{equation}\label{gaussianmuprior}
P(\mu  |\, z) = N[\mu \, | \,  f(z) ,  \sigma^2_\mu = [f'(z)]^2(\sigma^2_{\text{pec}}/c^2 + \sigma^2_z ) ].
\end{equation}
In the low-$z$ regime, where $d_\text{L}(z)$ is linear in $z$ (the Hubble law), the variance is 
\begin{equation}
\sigma^2_{\mu} =  \left(\frac{5}{z \ln 10} \right)^2 \left[ \sigma_z^2 + \frac{\sigma_{pec}^2}{c^2} \right].
\end{equation}

Under the assumption that peculiar velocity uncertainty amounts to Gaussian noise, at recession velocities $cz < 5 \sigma_{pec}$, the 
approximation of $P(\mu | z)$ with a normal distribution breaks down, due to the non-linearity of the logarithm.   This effect is 
inconsequential for our analysis because even though the distribution becomes non-Gaussian, at such low recession velocities, its 
width in magnitudes is much larger than the dispersion in SN Ia absolute magnitudes.  The redshift of a very low-$z$ SN Ia carries 
little information about the absolute magnitude, so that $P(\mu | z)$ is essentially flat over the width of the posterior density in $\mu$ 
conditioning on the light curves of the SN.  Hence, the exact form of  $P(\mu | z)$  is irrelevant in this regime.  SN Ia light curves can be 
used to infer the distances of these near-field supernovae and to measure the the local velocity field \citep{riess95,haugbolle07, 
neill07}.

\subsubsection{The SN Ia Population Distribution}

The utility of Type Ia SN for cosmological studies lies in the observed correlation of their peak luminosities with the shapes of their 
optical light curves.   Although the peak optical luminosities of SN Ia range over a factor of 3, using correlations with light curve shape 
reduces the scatter about the Hubble line to less than $\sim 0.20$ mag.   Physical modeling and simulation of SN Ia progenitors may 
provide useful explanations for the observed relationships between the observable properties of SN Ia and their  peak luminosities.   
Such work may also describe detailed probabilistic relationships between the two.  For example, we can define the joint population 
distribution of absolute magnitudes (at the reference time) and the observable light curve shapes in multiple passbands.  In our 
notation this population distribution is 
$P( \bm{\psi} | \, \text{Physical Parameters} )$
where the ``Physical Parameters'' may include, for example, the mass of the progenitor, the chemical composition and distribution 
within the progenitor, and the details of the explosion mechanism.  \citet{hillebrandt00} provide a review of progress in SN Ia explosion 
modeling.

In the absence of such detailed information, we learn the probabilistic relationships from the data.  We describe the SN Ia 
population distribution as $P( \bm{\psi} | \, \bm{\mu}_\psi, \bm{\Sigma}_\psi)$.  Here, $\bm{\mu}_\psi$ is a vector of hyperparameters 
that describe the average intrinsic characteristics, and $\bm{\Sigma}_\psi$ is a collection of hyperparameters describing the dispersion 
(variance) and correlations of the intrinsic characteristics of absolute light curves.   Since we have no information on these 
hyperparameters \emph{a priori}, we seek to estimate them (and their uncertainties) from the data.  

We will include this distribution in the global posterior density in the mathematical form of a ``prior'' on the intrinsic parameters $
\bm{\psi}$ of a single SN.  However, it is better to think  of this distribution as a ``population''  distribution from which the intrinsic 
parameters are randomly drawn.  Its hyperparameters are unknown and must be estimated simultaneously from the data.   It has a different 
interpretation in the context of the hierarchical model than the fixed prior of the simple Bayesian treatment (which has no hyperparameters 
to be estimated).

Since we have no \emph{a priori} information on the functional form of $P( \cdot | \, \bm{\mu}_\psi, \bm{\Sigma}_\psi)$, we must make 
some assumptions.  The simplest choice for a multivariate probability density that models correlations between parameters is the 
multivariate Gaussian.   In the rest of this paper we will assume $P( \cdot | \, \bm{\mu}_\psi, \bm{\Sigma}_\psi)  = N( \cdot | \, \bm{\mu}_
\psi, \bm{\Sigma}_\psi)$ with an unknown mean vector $\bm{\mu}_\psi = \mathbb{E}(\bm{\psi}_s)$ and unknown covariance matrix $
\bm{\Sigma}_\psi = \text{Var}(\bm{\psi}_s)$.  The intrinsic parameters of individual supernovae are independent, identically distributed 
random variables drawn from this probability density: 
$\bm{\psi} \sim N( \bm{\mu}_\psi, \bm{\Sigma}_\psi).$
If  the data indicate a different distribution from the one we have assumed,  we can attempt another choice of the form of the intrinsic 
population distribution.

The population hyperparameters $\bm{\mu}_\psi, \bm{\Sigma}_\psi$ are the most important variables in this hierarchical framework.  
During the training process, they model the intrinsic statistical properties of the SN Ia light curves, including the average behavior, 
intrinsic variability, correlations between different modes of light curve shape variation, correlations between absolute magnitudes in 
different filters, and cross-correlations between light curve shape parameters and the absolute magnitudes.  When the model is 
used to make predictions, they are crucial for using this information, and its uncertainty, to make distance estimates for new SN Ia light 
curves.

\subsubsection{Incorporating Dust Information}

Dust along the line of sight from the supernova to the observer causes both extinction and reddening of the emitted light.   These 
effects originate from Galactic dust, which has been measured and mapped \citep{sfd98}, and dust in the supernova's host galaxy, 
which is often more important, poorly understood and is currently the largest systematic uncertainty in cosmological inference with SN 
Ia \citep{conley07}.   Previous efforts to estimate dust extinction from SN Ia color excesses include \citet{riess96b}, \citet{phillips99}, 
and \citet{krisciunas00}.

We incorporate the  effects of host galaxy dust on supernova observations probabilistically within the full statistical model as follows.  
The extinction in a given passband is denoted $A_F$.  Assuming a CCM reddening law \citep{ccm89}, the extinction in a given band 
can be related to the visual extinction $A_V$ by $A_F/A_V = a_F + b_F R_V^{-1} $, so that there are two free parameters: the 
magnitude of visual extinction $A_V$ and the slope of the extinction law in the optical bands $R_V$.  The fixed regression coefficents 
$a_F$ and $b_F$ are determined from the dust reddening analysis of supernova spectra.
\citet{jha07} suggested using an exponential prior on the nonnegative   $A_V$ extinction to a particular supernova.   Interpreted as a 
population distribution, this can be incorporated into our framework as:
\begin{equation}\label{dustprior}
P(A_V, R_V | \tau_{A_V}, \bm{\alpha}_R) = \text{Expon}( A_V | \tau_{A_V}) P(R_V | \, \bm{\alpha}_R)
\end{equation}
where $\tau_{A_V}$ is a hyperparameter describing the exponential scale length or the average amount of visual extinction to the 
population of SN Ia.  The form of the population distribution of the $R_V$ is unknown, but we suppose it may be specified with 
hyperparameters $\bm{\alpha}_R$.  For example, if $R_V$ is fixed to a single known value, e.g. $\alpha_R = 1.7$ or $3.1$, we may 
set $P(R_V | \alpha_R) = \delta(R_V - \alpha_R)$.     Or one may allow the $R_V$ to vary within a population probability density, e.g. 
$R_V \sim N(\mu_R, \sigma^2_R)$, where the hyperparameters $\bm{\alpha}_R = (\mu_R, \sigma^2_R)$ may be fixed  or we may 
attempt to learn them from the data, if the data is sufficiently informative.  It is not known \emph{a priori} whether $A_V$ and $R_V$ can 
be treated as  independent random variables and if the population probability density of $(A_V, R_V)$ is separable as  indicated in Eq. 
\ref{dustprior}.

When modeling the near infrared observations, it makes more sense to reference the extinction values to  the $H$-band, rather than to 
$A_V$.  For given $A_H, R_V$ values, the extinction in any band, $A_F$ can be computed from the dust law.
The ratio of near infrared to visual extinction is roughly $A_H/A_V \sim 0.2$.
Since the behavior of dust in the near-infared is relatively insensitive to the slope of the reddening law in the optical bands, $R_V$,  it 
makes sense to set it to a fixed representative global value $\alpha_R$.  With this choice, the extinction population distribution is:
\begin{equation}
P(A_H, R_V | \tau_{A_H}, \alpha_R) = \text{Expon}( A_H | \tau_{A_H}) \delta(R_V - \alpha_R)
\end{equation}
where $\tau_{A_H}$ is the exponential scale length in magnitudes of the population distribution of $H$-band extinctions. To simplify 
the notation, we denote this hyperparameter as $\tau_A \equiv \tau_{A_H}$, which controls the dust scale for all filters through the reddening law.

\subsubsection{Conditionally Conjugate Hyperpriors}

To estimate the population hyperparameters $\bm{\mu}_\psi$ and $\bm{\Sigma}_\psi$, we make our priors on them explicit,  called 
\emph{hyperpriors}.  If we lack any external motivating evidence, we should choose non-informative or diffuse hyperpriors.  
Additionally, it is convenient to choose the hyperprior from a parametric family that is conditionally conjugate to the parametric family of 
the population density.  This means that the posterior density of the hyperparameters conditioned on the values of the other 
parameters and data is from the same parametric family of probability densities as the hyperprior.  This property is advantageous 
because if one can sample directly (generate random numbers from) the hyperprior density, then one can sample directly from the 
conditional posterior density of the hyperparameters.  This is useful for constructing Markov chains for statistical 
computation of the posterior density using Gibbs sampling (section \ref{statcomp}).

The hyperparameters  $\bm{\mu}_\psi$ and $\bm{\Sigma}_\psi$ describe a multivariate normal density on $\bm{\psi}_s$.  The 
conjugate family to the multivariate normal with unknown mean and covariance matrix is the Normal-Inverse-Wishart.  This hyperprior 
can be expressed as $P(\bm{\mu}_\psi, \bm{\Sigma}_\psi) = P(\bm{\mu}_\psi | \, \bm{\Sigma}_\psi) P(\bm{\Sigma}_\psi)$ such that
\begin{equation}\label{hyperprior}
\bm{\Sigma}_\psi \sim \text{Inv-Wishart}_{\nu_0}(\bm{\Lambda}_0^{-1})
\end{equation}
\begin{equation}\label{hyperprior2}
\bm{\mu}_\psi | \, \bm{\Sigma}_\psi \sim N(\bm{\mu}_0, \bm{\Sigma}_\psi / \kappa_0).
\end{equation}
The non-informative or diffuse conditionally conjugate density is obtained in the limit as $\kappa_0 \rightarrow 0$, $\nu_0 \rightarrow 
-1$, $| \bm{\Lambda}_0 | \rightarrow 0$ \citep{gelman_bda}  (the conventions regarding the Wishart distributions differ:  we 
choose the convention that if $\bm{W} \sim \text{Inv-Wishart}_\nu(\bm{S}^{-1})$ is a random matrix, then $\mathbb{E}(W) = \bm{S}/(\nu-d-1)$, 
where $d$ is the dimension of the $d \times d$ covariance matrix).  In this limit, the hyperprior of the population mean $\bm{\mu}_\psi$ 
becomes flat and the hyperprior of the  covariance $\bm{\Sigma}_\psi$ is a diffuse distribution over the space of positive semi-definite 
matrices, so that the hyperprior does not favor any particular solution.

For the extinction exponential scale length $\tau_A > 0$, we choose a uniform positive hyperprior, expressing no prior preference for a 
particular value.  If this is viewed as an  $\text{Inv-Gamma}(-1, 0)$ density on $\tau_A$, it is conditionally conjugate to the exponential 
distribution.

\subsubsection{The Global Posterior Density}

We now have the elements necessary to construct the full joint posterior density of the sample of SN Ia.  A single SN $s$ with multi-
band light curve data $\mathcal{D}_s = \{\mathcal{D}^F_s\}$, and redshift $z_s$, is described by intrinsic light curve parameters $
\bm{\psi}_s$, observable parameters $\bm{\phi}_s$, and distance modulus $\mu_s$, with dust extinction modeled by $A_H^s$ and 
$R_V^s$.
The relations between these parameters can be encoded as follows.  Let $\bm{v}$ be a constant indicator vector with the $j$th 
component $v^j$ defined as:
\begin{equation}
v^j \equiv \begin{cases} 1, & \text{if } \phi^j, \psi^j \text{ are magnitudes, e.g. } F_0 \text{ or } M_F\\
0, & \text{if } \phi^j, \psi^j \text{ are shape parameters.}\end{cases}
\end{equation}
Furthermore, define the vectors $\bm{A}_s$ with $j$th component $A^j_s$:  if $\phi_s^j$ and $\psi_s^j$  are magnitudes in band  $F$, 
then
\begin{equation}
A^j_s \equiv  A_{F,s}^\text{Gal} + A_F(A_H^s, R_V^s),
\end{equation}
otherwise, $A^j_s = 0$ if  $\phi_s^j$ and  $\psi_s^j$ are shape parameters.
The non-zero components depend on the host galaxy reddening law and $H$-band extinction, $A_{F,s}^\text{Gal}$ is the Galactic 
extinction, and $A_F(A_H^s, R_V^s)$ is the dust extinction in filter $F$ as a function of $A_H^s$ and $R_V^s$ using the dust law.  The 
relationship between the intrinsic and observable parameters of supernova $s$ can then be written compactly as:
\begin{equation}
\bm{\phi}_s = \bm{\psi}_s + \bm{v} \mu_s + \bm{A}_s.
\end{equation}
This equation encodes the relationship between apparent magnitudes, absolute magnitudes, extinction and distance moduli.  
In this expression, neither dust nor distance modify the light curve shape parameters, $\bm{\theta}^F_s$ common to both the 
observable $\bm{\phi}_s$ and intrinsic $\bm{\psi}_s$ vectors.
The joint posterior probability density for the parameters of a single supernova, conditioned on the values of the hyperparameters and 
the data, is proportional to the product of
\begin{itemize}
\item the probability of observing the photometric data given the apparent light curve, 
\item the probability of the distance modulus given the measured redshift, 
\item the probability of an absolute light curve equal to the apparent light curve minus the distance modulus and extinction, and 
\item the probability of the extinction value and dust law, 
\end{itemize}
conditioned on the population hyperparameters of the absolute light curves and dust properties:
\begin{equation}\label{singlesn}
\begin{split}
P&( \bm{\phi}_s , \mu_s, A_H^s, R_V^s | \, \mathcal{D}_s, z_s;  \bm{\mu}_\psi, \bm{\Sigma}_\psi, \tau_A, \alpha_R) \\ &\propto 
P( \mathcal{D}_s | \bm{\phi}_s) \times P(\mu_s | \, z_s) \\
&\times P(\bm{\psi}_s =  \bm{\phi}_s  - \bm{v} \mu_s - \bm{A}_s | \, \bm{\mu}_\psi, \bm{\Sigma}_\psi) \\ &\times P(A_H^s, R_V^s |\, \tau_A, 
\alpha_R ) .
\end{split}
\end{equation}
Now consider the full database of SN Ia light curves $\mathcal{D} = \{ \mathcal{D}_s \}$ with measured cosmological redshifts $
\mathcal{Z} = \{ z_s \}$.  The global joint posterior density of all supernova observables $\{ \bm{\phi}_s \}$, distance moduli $\{\mu_s\}$, 
dust parameters $\{A_H^s, R_V^s\}$ and the population hyperparameters conditioned on the database $\mathcal{D}, \mathcal{Z}$  is proportional to 
the product of $N_{\text{SN}}$ individual conditional posterior densities multiplied by the hyperpriors.
\begin{equation}\label{globalposterior}
\begin{split}
P&(\{ \bm{\phi}_s , \mu_s, A_H^s, R_V^s\} ;  \bm{\mu}_\psi, \bm{\Sigma}_\psi, \tau_A, \alpha_R | \, \mathcal{D}, \mathcal{Z}) \\
&\propto \left[ \prod_{s=1}^{N_{\text{SN}}} P( \bm{\phi}_s , \mu_s, A_H^s, R_V^s | \, \mathcal{D}_s, z_s;  \bm{\mu}_\psi, \bm{\Sigma}_\psi, 
\tau_A, \alpha_R) \right] \\
&\times P(\bm{\mu}_\psi, \bm{\Sigma}_\psi) \times P(\tau_A, \alpha_R)
\end{split}
\end{equation}
For full generality we have derived the global joint posterior density in the case that we wish to estimate the probable values of the 
hyperparameters $\alpha_R$ from the data. If we  fix the $R_V^s$ to a fixed global value  $\alpha_R$, this is equivalent to evaluating 
the above joint density conditioned on $R_V^s = R_V = \alpha_R$.  All fully Bayesian inferences on the remaining parameters and 
hyperparameters are based on mapping out this global joint posterior density.  The marginal posterior density of the hyperparameters, 
$P(\bm{\mu}_\psi, \bm{\Sigma}_\psi, \tau_A | \, \mathcal{D}, \mathcal{Z}, R_V)$ or $P(\bm{\mu}_\psi, \bm{\Sigma}_\psi, \tau_A, \alpha_R 
| \, \mathcal{D}, \mathcal{Z})$ is obtained by integration over the individual SN parameters $\{ \bm{\phi}_s , \mu_s, A_H^s, R_V^s\}$ or 
$\{ \bm{\phi}_s , \mu_s, A_H^s\}$.

We refer to the set of SN in $\mathcal{D}, \mathcal{Z}$ as a training set,  and ``training'' means  computing the marginal posterior density of 
the hyperparameters, conditioned on this data.   In the Bayesian paradigm we are interested not just in point estimates of the 
hyperparameters, e.g. ``best values''  $\hat{\bm{\mu}}_\psi, \hat{\bm{\Sigma}}_\psi, \hat{\tau_A}$, but on their joint posterior probability 
density as a quantification of their uncertainties.

\subsubsection{The Predictive Posterior Density}\label{prediction}

The ultimate purpose of SN Ia light curve inference is to estimate luminosity distances to  distant SN that are not included in the nearby, 
low-$z$ training set. That is, given observations of a new supernova's multi-band light curve $\tilde{\mathcal{D}}_s$, we wish to fit the 
light curve model and to predict the distance modulus.  Prediction differs from training in the fact that we do not use any prior 
information on the distance (from e.g. the redshift) in our probability calculus.  The predictive posterior density for the new supernova $
\tilde{s}$ (with parameters denoted by tilde) conditioned on the population hyperparameters and the new light curve data $
\tilde{\mathcal{D}}_s$ is: 
\begin{equation}
\begin{split}
P&( \tilde{\bm{\phi}}_s , \tilde{\mu}_s, \tilde{A}_H^s, \tilde{R}_V^s | \, \tilde{\mathcal{D}}_s; \bm{\mu}_\psi, \bm{\Sigma}_\psi, \tau_A, 
\alpha_R) \\
&\propto P( \tilde{\mathcal{D}}_s | \tilde{\bm{\phi}}_s)  \\
&\times P(\tilde{\bm{\psi}}_s =  \tilde{\bm{\phi}}_s  - \bm{v} \tilde{\mu}_s - \tilde{\bm{A}}_s | \, \bm{\mu}_\psi, \bm{\Sigma}_\psi) \\ &\times 
P(\tilde{A}_H^s, \tilde{R}_V^s |\, \tau_A, \alpha_R ) .
\end{split}
\end{equation}
We must also incorporate our (joint) uncertainties of the hyperparameters.  This is encapsulated in the marginal posterior density of the 
hyperparameters from the training set. The full predictive posterior probability density for the new supernova $\tilde{s}$ is the previous 
expression multiplied by the training posterior density $P(\bm{\mu}_\psi, \bm{\Sigma}_\psi, \tau_A, \alpha_R | \, \mathcal{D}, 
\mathcal{Z}) $ and integrated over the probability of the hyperparameters  $\bm{\mu}_\psi, \bm{\Sigma}_\psi, \tau_A, \alpha_R$.
The marginal predictive posterior density of the new supernova's distance modulus $\tilde{\mu}_s$, $P( \tilde{\mu}_s | \, 
\tilde{\mathcal{D}}_s, \mathcal{D}, \mathcal{Z})$, is obtained by integrating this over the remaining parameters, $
\tilde{\bm{\phi}}_s , \tilde{A}_H^s, \tilde{R}_V^s$.

\subsection{Representation as a Directed Acyclic Graph}\label{dag_section}

We have constructed the posterior density of all individual parameters and population-level hyperparameters conditioned on the 
observed dataset of multi-band SN Ia light curves.  This was done by layering relationships of conditional probability.   All hierarchical 
joint probability densities of data and parameters can be represented in terms of a probabilistic graphical model known as a directed 
acyclic graph.  The graph consists of nodes representing parameters and data connected by arrows that represent probabilistic 
dependencies.  It obeys the restriction that there are no \emph{directed cycles}, i.e. it is impossible to move from any node along the 
arrows and return to the same node.  The acyclic requirement ensures that inference from the posterior density  contains no loops of 
circular logic. It is useful to
represent complex inference problems, involving many potential
sources of randomness, with an equivalent directed acyclic
graphical model.  Although all the information about the model 
and data is expressed by writing down the joint probability density explicitly,  probabilistic graphical models  serve as a useful visual 
representation of the structure of the hierarchical model and their interface with data.
Formal graphical models have not been used before in SN Ia inference, and they are not prevalent in astronomy, so we provide a 
basic introduction below.  Further background and theory of graphical models can be found in \citet{bishop}, \citet{jensen01} and 
\citet{pearl88}.

The directed graph is constructed as follows.  Each parameter or datum corresponds to a node (or vertex).  Conditional relationships 
between nodes are encoded using directed links or arrows (edges).   Hence the joint probability of two variables $P(x,y) = P(x) P(y | x)
$ is represented by $x \rightarrow y$.  For obvious reasons, the parameter $x$ is termed the \emph{parent} and $y$ is termed the 
\emph{child}.  More generally, in a high-dimensional problem, if  there exists a directed path of any length between node $x$ and 
another node $y$, then $y$ is a \emph{descendant} of $x$.  The joint probability distribution over $N$ random variables $\theta_i$ 
represented by a directed graph can be written as the product of the conditional probabilities (the factorization):
\begin{equation}\label{factorization}
P(\{ \theta_i \}) = \prod_{i=1}^N P( \theta_i | \{ \text{Parents of } \theta_i \})
\end{equation}

If a parameter is observed (and thus conditioned on), its node is shaded.  If the parameter is unknown and hidden, it is left open.   The 
graph clearly distinguishes between the observed data and the hidden variables that are inferred.  Graphical models are most useful 
in inference problems involving many potentially interacting components or sources of randomness.  The complexity of the problem is 
reflected in the connectedness of the graph.   The links between the nodes encode statements of conditional independence (\S \ref{dsep}).

\subsubsection{Directed Graph for Model Training}

\begin{figure}[t]
\centering
\includegraphics[angle=0,scale=0.4]{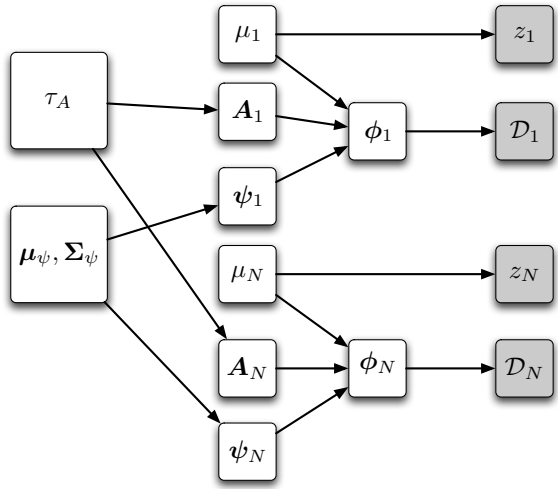}
\caption{\label{dag}Directed acyclic graph for hierarchical Bayesian inference from a training set of Type Ia SN light curves.  This is a 
graphical representation of the joint distribution of unknown parameters and observations for a training set of $N$ SN Ia.  Each 
parameter is represented by a node, and the links between node indicate relationships of conditional probability.  The variables in the 
far left column are the hyperparameters which describe the population probability distribution of supernova characteristics, and the 
population distribution of extinction values.  The variables in the middle left column describe the distances, extinctions, and absolute 
light curves of individual supernovae.  The variables in the middle right column are the observable parameters that describe the 
apparent light curves of individual SN Ia.  The final column contains the observations of the redshifts and multi-band light curves of 
individual SN Ia.  The open nodes describe unknown and hidden parameters, whereas the shaded nodes describe observed values 
that are conditioned upon in the posterior density.}
\end{figure}

The directed graph corresponding to the global posterior density conditioned on the training set $\mathcal{D}$ of $N$ SN Ia is shown 
in Figure \ref{dag}.   The pathways from the roots to the data can be understood as a generative model for  a data set of SN Ia light 
curves.   At the highest hierarchical level (far left), the population distributions for SN Ia and dust extinction are described by unknown 
hyperparameters $\bm{\mu}_\psi, \bm{\Sigma}_\psi$, and $\tau_A$.   At the next level, each supernova $s$ draws intrinsic light curves 
$\bm{\psi}_s$ and extinction values $\bm{A}_s$ from these populations as  independently and identically distributed random 
samples.  These hidden parameters combine with the supernova distance modulus $\mu_s$ to produce the observable light curve $
\bm{\phi}_s$, which is sampled with measurement noise to produce the photometric light curve data $\mathcal{D}_s$.  In addition, the 
(hidden) distance modulus is associated with a redshift $z_s$, observed with both measurement noise and peculiar velocity 
uncertainty.  For a sample of SN Ia light curves, this process is replicated $N_{\text{SN}}$ times.  Our goal is to compute the joint 
posterior density of all the open nodes conditioned on the data in the shaded nodes.


The conditional independence properties of the graph and model imply that, although the individual parameters of one SN are \emph{conditionally} 
independent from those of a different SN, given the population hyperparameters, they are not \emph{marginally} independent.  The 
hidden hyperparameters are unknown \emph{a priori};  they must be learned from the data jointly with the individual SN parameters.   
Thus, the full graph does not factor into independent $N_\text{SN}$ subgraphs.  We must condition the whole graph and the global 
joint density on a database of many SN Ia light curves simultaneously rather than on each supernova individually.

Figure \ref{dag} shows that the SN Ia population hyperparameters $\bm{\mu}_\psi, \bm{\Sigma}_\psi
$ are conditionally independent from every other parameter and datum in the graph, given the intrinsic SN parameters $\{\bm{\psi}_s \}
$: $P(\bm{\mu}_\psi, \bm{\Sigma}_\psi | \, \cdot, \mathcal{D}, \mathcal{Z}, \{\bm{\psi}_s \})  = P(\bm{\mu}_\psi, \bm{\Sigma}_\psi | \{\bm{\psi}_s \})$.     Here we use $(\cdot)$ to indicate all the other parameters in the global joint density that have not been denoted 
explicitly.  
 Similarly, given all the extinction values of the supernovae, $ \{ \bm{A}_s \}$,  the extinction population exponential scale is 
conditionally independent from all other parameters (and data), so that  $P(\tau_A | \, \cdot, \mathcal{D}, \mathcal{Z}, \{ \bm{A}_s \}) = 
P( \tau_A |\, \{ \bm{A}_s \})$.  


The graph also shows that $\bm{\psi}_s$, $\mu_s$ and $\bm{A}_s$ are conditionally dependent in the posterior distribution, because their 
descendant $\mathcal{D}_s$ is observed, even though they are \emph{a priori} independent random variables.  This dependency is 
reflects the tradeoffs involved in explaining the observed light curves as a combination of random fluctuations due to dust, intrinsic 
randomness of the absolute light curves, and distance uncertainties attributed to peculiar velocities.  The Bayesian approach is not to 
pick out just one possible combination of the separate factors, but to consider the probability distribution over the whole ensemble of 
hypotheses.


Another consequence of this conditional dependency is that there are unblocked paths between the SN Ia population 
hyperparameters,  $\bm{\mu}_\psi$ and $\bm{\Sigma}_\psi$ and the dust extinction hyperparameter $\tau_A$.  These paths pass 
through the conditionally dependent parameters $\bm{A}_s$, $\bm{\psi}_s$, and $\bm{\phi}_s$ for each supernova.  Thus, the 
population hyperparameters are also conditionally dependent.  This implies that posterior inferences of  $\bm{\mu}_\psi, \bm{\Sigma}_
\psi$ and those of $\tau_A$ cannot be separated. This is why we take the global approach, conditioning the global posterior density on 
the entire data set simultaneously, and exploring the complete joint parameter space.

The conditional independence structure implied by the graph depends neither on the  choices of distributions made in Section 
\ref{constructposterior}, nor on the particular functional light curve model that is assumed.   We depicted the directed graph for 
inference with fixed $R_V$.   If we wish to learn about $R_V$, it would become a random variable with a population distribution.  
Hence the graph would include nodes for each $R_V^s$ and a node for the hyperparameters $\alpha_R$, with the appropriate links.

\subsubsection{Directed Graph for Prediction}

\begin{figure}[t]
\centering
\includegraphics[angle=0,scale=0.4]{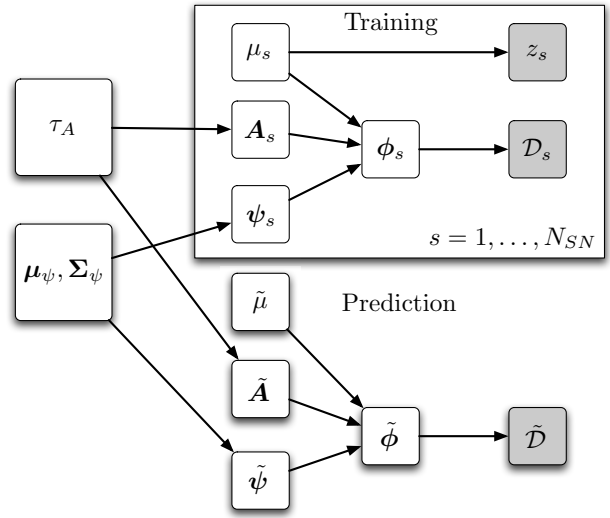}
\caption{\label{dag_pred}Directed acyclic graph for training and prediction with Type Ia SN light curves.  The rectangle depicts a plate 
representing the $N_{\text{SN}}$ SN Ia in the training set.  The tilde parameters describe a new supernova for which we seek to predict 
the distance modulus. The open nodes describe unknown and hidden parameters, whereas the shaded nodes describe observed 
values that are conditioned upon in the predictive posterior density.}
\end{figure}

The directed graph for the prediction task using data from a new supernova is presented in Figure \ref{dag_pred}.   We 
depict the entire training set of supernovae on a \emph{plate} which is understood to represent $N_{\text{SN}}$ different instances.  
The quantities relevant to the prediction supernova are labeled with tildes.  The essential difference between training and prediction is 
that in the training set we use distance information from the redshift, whereas in prediction we do not.   The task of prediction is to infer 
the joint probability density of the hidden quantities $\tilde{\mu}$, $\tilde{\bm{A}}$, and $\tilde{\bm{\psi}}$ by fitting the light curve data $
\tilde{\mathcal{D}}$ described by the observable parameters $\tilde{\bm{\phi}}$ plus measurement noise.  The unblocked paths 
between the training set and the prediction set depict how information from the training set constrains the population hyperparameters 
(i.e. by informing the posterior density), which in turn pass that information (and its uncertainty) onto the prediction variables.  The 
marginal predictive posterior density for the new supernova's distance modulus is obtained by integrating over the uncertainties in the 
population hyperparameters $\bm{\mu}_\psi$, $\bm{\Sigma}_\psi$, and $\tau_A$, and over the extinction $\tilde{\bm{A}}$, magnitudes 
and the shape parameters, $\tilde{\bm{\phi}}$.

\subsection{Statistical Computation of the Global Posterior Density}\label{statcomp}

The global posterior probability density of all parameters and hyperparameters of the full model conditioned on the training set  
database of SN Ia observations, Eq. \ref{globalposterior}, is a function of many variables.  Consider a minimal model that does not 
account for dust extinction.  We suppose it has one shape parameter $\theta$, and models light curves in three filters.  There are four 
observable parameters, plus one for the distance modulus, for each supernova.  In addition, the hyperparameters $\bm{\mu}_\psi$ and 
$\bm{\Sigma}_\psi$ contain four plus ten variables (since the covariance matrix of the absolute magnitudes and light curve shape 
parameters must be symmetric).  Suppose a minimal training set contained observations of forty SN Ia light curves in the three filters.  
The total number of variables, which is the dimensionality of the space over which the global posterior density is defined, is 214.  
Clearly, mapping the joint posterior on a rectangular multidimensional grid is intractable.  Even a relatively crude grid, with only five 
points per dimension, would require more than $10^{149}$ evaluations of the posterior density.  

To address the complexities of hierarchical inference with realistic data sets, it would be practically useful to construct a statistical 
inference approach without appealing to the asymptotic results from large-sample theory.  The only way to account for all uncertainties 
in the model parameters consistently is to compute the full hierarchical joint density, Eq. \ref{globalposterior}, and estimate all 
individual parameters and hyperparameters simultaneously, conditioned on the entire SN Ia database.  Marginal estimates of 
parameters are obtained by integration over non-Gaussian uncertainties.  However, the obstacles to this approach are two-fold:   (1) 
we must compute the global posterior density in a parameter space with hundreds of dimensions, and (2) the marginal estimates of 
single parameters require integration of the posterior density over hundreds of other parameters.

We tackle both of these problems by using stochastic simulation techniques to sample the full parameter space efficiently.  In this 
section we describe the construction and operation of a Markov Chain Monte Carlo (MCMC) algorithm that takes advantage of the 
conditional independence structure evident in the directed acyclic graph (Fig. \ref{dag}) of Section \ref{dag_section} to sample the 
global posterior density, Eq. \ref{globalposterior}.

\subsubsection{Metropolis-Hastings and the Gibbs Sampler}

Markov Chain Monte Carlo is a general and well-established technique for statistical analysis and is well suited for Bayesian 
computations.   It is employed, for example, in CMB and joint cosmology analyses \citep{lewis02,tegmark04}, for fitting light curve 
models \citep{mandel02} to planetary transit observations \citep{holman06}, and for radial velocity analysis of extrasolar planetary 
systems \citep{ford05}.
Since MCMC has not been used previously in SN Ia light curve inference methods, we briefly review some basic elements of MCMC to 
establish terminology.  More thorough treatments of MCMC methods and theory can be found elsewhere \citep{gilks95, 
liu02,gelman_bda}.  

The purpose of an MCMC algorithm is to generate a Markov chain stochastic process that is irreducible and ergodic, and converges in 
probability to a stationary distribution that is the same as the target distribution (the posterior density).   Upon convergence, the 
probability that the chain is in a particular state is equal to the posterior density of the state, and the proportion of time the chain spends 
in a given region of parameter space is proportional to the posterior probability of that region.  Hence, MCMC can be used to generate 
many samples from an arbitrary, complex probability distribution (which cannot be sampled from directly), and those samples can be 
used to represent the target distribution and compute characteristics of the distribution, such as means, modes, intervals and integrals.  

The cornerstone of many MCMC implementations is the Metropolis-Hastings algorithm.  
Suppose our target posterior probability density is $P(\bm{\theta} | \mathcal{D} )$ for a vector of generic parameters $\bm{\theta}$, and 
can be computed up to a normalization constant.  The MCMC algorithm  generates a sequence of samples.  Let $\bm{\theta}^t$ denote 
the $t$th sample.  We start with some initial estimate $\bm{\theta}^{t=1}$, and generate subsequent values of the chain as follows.  We 
select a proposal (or jumping) probability density $Q(\bm{\theta}^* | \bm{\theta})$, giving  the probability of proposing $\bm{\theta}^*$ 
for the next value given that the current state is $\bm{\theta}$.  This proposal density is chosen so that it can be directly sampled (e.g. a 
Gaussian).  If the current state is $\bm{\theta}^t$, then we generate a proposal $\bm{\theta}^*$ from $Q(\bm{\theta}^* | \bm{\theta}^t)$.  
We then compute the Metropolis-Hastings ratio:
\begin{equation}\label{mhratio}
r = \frac{P(\bm{\theta}^* | \mathcal{D}) / Q(\bm{\theta}^* | \bm{\theta}^t)}{P(\bm{\theta} | \mathcal{D}) / Q(\bm{\theta}^t | \bm{\theta}^*)}
\end{equation}
The proposal $\bm{\theta}^*$ is accepted ($\bm{\theta}^{t+1} = \bm{\theta}^*$) with probability $\min(r,1)$.   If it is not accepted, the 
proposal is rejected and the next value of the chain is the same as the current value $\bm{\theta}^{t+1} = \bm{\theta}^t$.  In the next 
iteration a new proposal is generated from $Q(\bm{\theta}^* | \bm{\theta}^{t+1})$ and the algorithm repeats.  

A special case of Metropolis-Hastings is the random-walk Metropolis algorithm in which the proposal distribution is symmetric 
$Q(\bm{\theta}^* | \bm{\theta}) = Q(\bm{\theta} | \bm{\theta}^*)$ and the proposal is centered around the current position, e.g.  $
\bm{\theta}^* \sim N( \bm{\theta} , \sigma^2 \bm{I})$.  Gibbs sampling is  another very useful case of the Metropolis-Hastings rule and   
proceeds by simply drawing from the conditional probability of each block of parameters in turn, conditioning on the others as fixed, 
until all the parameters have been sampled.    We can employ Gibbs sampling in our hierarchical SN Ia framework because 
the model is built up from conditional relations, and many of the conditional posterior distributions can be directly sampled.   Our 
\textsc{BayeSN} algorithm uses a combination of these strategies to generate  efficient MCMC chains.

\subsubsection{The \textsc{BayeSN} Algorithm - Training }\label{bayesntrain}

We describe the \textsc{BayeSN} MCMC algorithm in the context of computing the global posterior in Eq. \ref{globalposterior} for fixed 
$R_V$.  Let 
$\mathcal{S} = ( \{\bm{\phi}_s,  \mu_s, A_H^s\} , \bm{\mu}_\psi, \bm{\Sigma}_\psi, \tau_A)$ 
 be a vector containing all the current values of the parameters and hyperparameters in the model (the ``state'' or position).  
\textsc{BayeSN} utilizes a sequential Gibbs sampling structure that updates blocks of parameters in $\mathcal{S}$.  After a full scan 
(after all parameters have been given a chance to update), the current state of $\mathcal{S}$ is recorded as an MCMC sample.  To 
begin a chain we populate $\mathcal{S}$ with a set of initial positions.   The \textsc{BayeSN} MCMC algorithm works in two stages: a) 
sampling the population hyperparameters conditional on the individual parameters, and b) sampling the individual supernova 
parameters conditional on the population hyperparameters. An outline of the Gibbs scan follows; more details are presented in the 
appendices \S\ref{bayesnmath}, \S\ref{practical}.

1. Update the SN Ia population hyperparameters, $\bm{\mu}_\psi, \bm{\Sigma}_\psi$, conditional on the current values $\{ \bm{\psi}_s\}
$, obtained from the current parameters: $\bm{\psi}_s = \bm{\phi}_s - \bm{v} \mu_s - \bm{A}_s$.  This is done by Gibbs sampling 
directly from the conditional posterior density $P( \bm{\mu}_\psi, \bm{\Sigma}_\psi | \, \cdot, \mathcal{D}, \mathcal{Z}) = P(  \bm{\mu}_
\psi, \bm{\Sigma}_\psi | \, \{ \bm{\psi}_s\})$.

2. Gibbs sample the extinction population hyperparameter $\tau_A$ from the conditional density $P(\tau_A | \cdot, \mathcal{D}, 
\mathcal{Z} ) = P(\tau_A | \{ A_H^s \})$.

Next we update the individual supernova parameters, conditional on the population hyperparameters we have just sampled.  The 
individual parameters of one supernova are conditionally independent from those of another supernova, given the hyperparameters.  
We cycle $s$ through the list of SN Ia, and for each supernova $s$ we repeat steps 3a to 3c to update observable parameters for each 
passband $F$.  Let $\bm{\phi}_s^{-F_0}$, $\bm{\phi}_s^{-\text{L},F}$, and  $\bm{\phi}_s^{-\text{NL},F}$ denote all the observable 
parameters in $\bm{\phi}_s$ other than the apparent magnitude, linear shape parameters, and nonlinear shape parameters in $F$, 
respectively. 

3a.  Gibbs sample the apparent magnitude $F_{0,s}$ by drawing directly from the conditional posterior $P(F_{0,s} | \, \bm{\phi}_s^{-
F_0}, \mu_s, \bm{A}_s; \bm{\mu}_\psi, \bm{\Sigma}_\psi, \tau_A, \mathcal{D}_s, z_s)$.

3b.  Gibbs sample the linear shape parameters $\bm{\theta}^F_{\text{L},s}$ in filter $F$ by drawing directly from
the conditional density,  $P(\bm{\theta}^F_{\text{L},s} | \, \bm{\phi}_s^{-\text{L},F}, \mu_s, \bm{A}_s; \bm{\mu}_\psi, \bm{\Sigma}_\psi, 
\tau_A, \mathcal{D}_s, z_s)$.

3c.  Random-walk Metropolis update the nonlinear shape parameters in band $F$, $\bm{\theta}_{\text{NL},s}^F$, using a jumping 
kernel $\bm{\Sigma}_{\text{jump},s}^{\text{NL},F}$ to move through the conditional density  $P( \bm{\theta}_{\text{NL},s}^F  | \, 
\bm{\phi}_s^{-\text{NL},F}, \mu_s, \bm{A}_s; \bm{\mu}_\psi, \bm{\Sigma}_\psi, \mathcal{D}_s, z_s)$.  

3d. Update the distance modulus $\mu_s$ using Metropolis-Hastings.   We propose a new $\mu_s$ drawn from a Gaussian 
approximation to the conditional posterior density $P(\mu_s | \, \bm{\phi}_s, \bm{A}_s ; \bm{\mu}_\psi, \bm{\Sigma}_\psi, \tau_A, 
\mathcal{D}_s, \mathcal{Z}_s)$, and use Metropolis-Hastings rejection to correct for the approximation.

3e. Update the extinction $A_H^s$ using a random-walk Metropolis step along the conditional density $P( A_H^s | \, \bm{\phi}_s, 
\mu_s; \bm{\mu}_\psi, \bm{\Sigma}_\psi, \tau_A)$, with a jumping scale $\sigma^2_{\text{jump},s}$.

4. Steps 3a to 3e are repeated for all SN Ia in the data set.  After all parameters have been updated, we record the current state of $
\mathcal{S}$ as an MCMC sample, and return to step 1.  After we have iterated $n$ times we finish with a Markov chain $
\bm{\mathcal{S}} = ( \mathcal{S}_1, \ldots \mathcal{S}_t, \ldots \mathcal{S}_n)$.

\subsubsection{\textsc{BayeSN} Algorithm - Prediction}\label{bayesn_pred}

The prediction mode of \textsc{BayeSN} follows essentially the same algorithm.    We assume that the prediction set is sampled from 
the same population as the training set.  This could be false, for example, if the SN Ia  in the prediction set had extremely different 
observed light curves.   This would also be false if either observational selection effects or progenitor evolution caused a distant 
prediction set to sample a different portion of the SN Ia population, or a physically different population, that is not represented in the 
nearby training set.   Training and prediction actually can be conducted simultaneously in a single run of the Gibbs sampler.  The main 
distinction is we do not condition on the redshifts of the SN in the prediction set, i.e. the factor $P(\mu_s | z_s)$ would be replaced by 
$P(\mu_s) \propto 1 $ in step 3d above.  With this change, the \textsc{BayeSN} algorithm will generate inferences on the graphical model in 
Fig. \ref{dag_pred}, for both the training and prediction set SN simultaneously. 

In many cases, however, we may wish to train the model on the training set SN once, and store the posterior inferences of the 
population hyperparameters.   To make predictions for new SN, we would recall this information and repeatedly apply it to the new 
data, without updating the training posterior inferences.  We can do this by making two changes to the above algorithm.  The goal is to 
generate a Markov chain $\bm{\mathcal{S}}_P$ that samples the predictive posterior density.  We assume we have already done a 
training MCMC and have a chain $\bm{\mathcal{S}}$ that samples the training posterior density conditioned on the training set $
\mathcal{D}, \mathcal{Z}$.  Steps 1, 2 and 3d change to:

1P \& 2P.  Draw the population hyperparameters $\bm{\mu}_\psi$, $\bm{\Sigma}_\psi$ and $\tau_A$ from the marginal posterior 
training density $P(\bm{\mu}_\psi,  \bm{\Sigma}_\psi,\tau_A | \mathcal{D}, \mathcal{Z})$.  This is easily done by picking a random 
sample from the training chain $\bm{\mathcal{S}}$ and using the values of the hyperparameters in that sample.

3dP.  Gibbs sample the predictive $\mu_s$ from $P(\mu_s | \, \bm{\phi}_s, \bm{A}_s ; \bm{\mu}_\psi, \bm{\Sigma}_\psi, \tau_A, 
\mathcal{D}, \mathcal{Z})$, omitting the factor $P(\mu_s | z_s)$ since we do not condition on the redshift for prediction SN $s$.

With these steps the algorithm is run to build up a Markov chain $\bm{\mathcal{S}}_P$ of samples from the predictive posterior density.

\section{Constructing Template Models for Near Infrared SN Ia Light Curves}

To complete the statistical model we must specify functional models for the normalized light curves, $l^F(t; \bm{\theta}^F)$, such that $F_0 + 
l^F(t; \bm{\theta}^F)$ describes the apparent light curve.  The function  $l^F(t; \bm{\theta}^F)$ captures variations in the light curve shapes in the 
observed filters.   There is significant freedom in defining these functions and they will generally depend on the application.  For the 
remainder of this paper, we apply the hierarchical model described above to the SN Ia data in the $JHK_s$ NIR bands.
In this section, we describe our methods for generating empirical template light curve models in the $JHK_s$ bands.

In the $H$ and $K_s$ bands, we  model the light curves using maximum likelihood templates that assume the 
normalized light curves of all the SN are identical, at least between -10 and 20 days from maximum light in $B$.  In the $J$-band, 
where the PAIRITEL photometry is better, we construct a light curve model that captures the shape variations, 
specifically the timing and amplitudes of the initial decline, trough, second rise, second peak and subsequent decline.  This Flexible 
Light-curve Infrared Template (FLIRT) model is constructed over the range -10 to 60 days past maximum light in $B$ and depends on 
four light curve shape parameters to desribe the features of the $J$-band light curve.

\subsection{Maximum Likelihood Light Curve Templates}\label{MaxL}

The simplest possible light curve model assumes that the normalized  light curve in filter $F$ of all SN are identical.  The 
normalized light curve data from all SN are sampled with noise from an underlying light curve function of the rest frame phase $l^F(t)$.   
A strategy for describing this function is to construct a template that is defined by a set of knots $\{\bm{p}^F, \bm{\tau} \}$ and an 
interpolation method.  The knots are defined such that $l^F(t = \tau_i) = p_i^F$, and $l^F(t) = \bm{s}(t; \bm{\tau}) \cdot \bm{p}^F$, where 
the vector $\bm{s}(t; \bm{\tau})$ is defined by an interpolation method that is linear in the  knot values $\bm{p}^F$.  We choose a 
natural cubic spline rule which ensures smoothness in the template up to two continuous derivatives \citep{numrec}.  It is convenient to 
choose the $\tau_i$ to lie on a regular grid with spacing $\Delta \tau$ such that one of the $\tau_j = 0$ coincides with the reference 
time ($T_{Bmax}$), forcing the corresponding $p_j^F = 0$, so that $l^F(t = \tau_j = 0) =  p_j^F = 0$ for the normalized light curve.   

\begin{figure}[b]
\centering
\includegraphics[angle=0,scale=0.45]{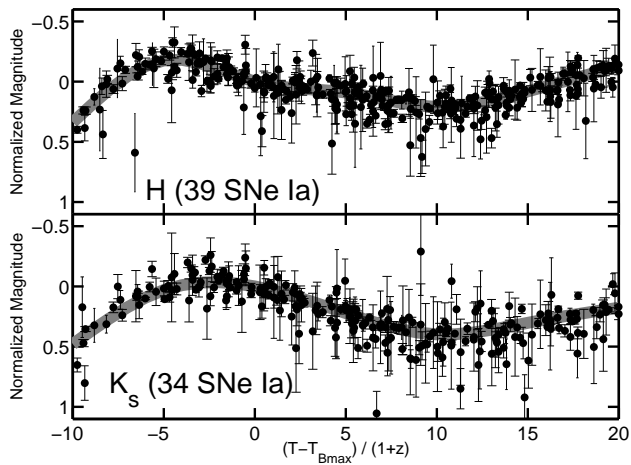}
\caption{\label{HKtemp}Maximum likelihood templates (grey curves) with the $H$ and $K_s$ band data.  The $H$ and $K_s$ 
normalized light curves of different SN are very similar between -10 and 20 days.}
\end{figure}

The $N_s^F$ photometric data points  from supernova $s$ are sampled with Gaussian noise from the model template plus a constant 
apparent magnitude offset $F_0^s$.   The joint posterior of the apparent magnitude offsets and the template for the data sets $
\mathcal{D}_s^F$ consisting of measurements from $N_{\text{SN}}$ supernovae in one band $F$ is proportional to the likelihood:
\begin{equation}
P( \{F_0^s\}, \bm{p}^F | \, \{\mathcal{D}_s^F \}) \propto\prod_{s=1}^{N_{\text{SN}}}  N( \bm{m}_s^F | \bm{1}F_0^s + \bm{S} \bm{p}^F , \, 
\bm{W}^F_s )
\end{equation}
where $\bm{S}$ is a matrix with $i$th row equal to $\bm{s}(t_i^s; \bm{\tau})$.
The joint maximum likelihood estimates of the apparent magnitudes $\{\hat{F}_0^s\}$ and the template $\hat{\bm{p}}^F$ is obtained by 
maximizing the log joint likelihood function subject to the linear constraint $f(0) = p_j = 0$.  This can be accomplished using the 
method of Lagrange multipliers.  The quadratic optimization problem can be solved easily  using non-iterative linear methods and 
yields a unique solution.  The maximum likelihood template light curve model produced this way is defined as $l^F(t)= \bm{s}(t; 
\bm{\tau}) \cdot \hat{\bm{p}}^F$ for band $F$. The $H$ and $K_s$ bands templates are depicted in Fig. \ref{HKtemp} and the values $
\bm{\hat{p}}^F$ are listed in Table 1.

\subsection{Flexible Light-curve InfraRed Template}\label{FLIRT}
\subsubsection{Definition}
Although the $J$-band light curves are very similar in shape near the peak, past 10 days after $B_{max}$, there are variations in the 
time and depth of the trough and the time and height of the second peak.
This behavior may be explained by the changes in the ionization states of iron in the SN atmosphere \citep{kasen06}.
We describe an empirical model for $J$-band light curves that transforms with simple scalings to mimic the individual decline 
timescales, trough depths, second rise timescales and second peak amplitudes of each supernova light curve.  These scalings are 
suggested by an inspection of the light curves.

We posit a $J$-band fiducial normalized light curve $f(
\lambda)$ that can be transformed to accommodate the variations in characteristic time scales and amplitudes observed in individual 
$J$-band light curves.  Here $\lambda$ is a feature time scale describing the timings of features in the fiducial light curve.   We can 
map this time scale to the chronological time scale $t$ of a particular supernova by introducing a time-warping function that allows the 
pre-trough phase to be scaled independently from the post-trough phase.  The rate at which the feature time maps to the chronological 
time is:
\begin{equation}
\frac{dt}{d\lambda} = \begin{cases} \alpha, & \text{if } \lambda \le \lambda_t ,\\
\beta, & \text{if } \lambda > \lambda_t \end{cases}
\end{equation}
where the parameters $\alpha$, $\beta$ are positive constants and of order one and $\lambda_t$ is the feature time of the trough in 
the fiducial light curve. The solution is $t(\lambda)  = \alpha \min(\lambda,\lambda_t) + \beta (\lambda-\lambda_t)^+$ where $u^+ = 
\max(u,0)$.   This function can be inverted as
\begin{equation}\label{invtw}
\lambda(t)  = \begin{cases} \alpha^{-1} t, & \text{if } t \le \alpha \lambda_t \\
\beta^{-1} t + \lambda_t (1-\alpha/\beta), & \text{if } t > \alpha \lambda_t. \end{cases}
\end{equation}
These equations represent simple transformations between the chronological and feature time axes, or the ``horizontal'' dimensions.   

Even after adjusting for variations in the two timescales, there are still variations in the depth of the trough and the amplitude of the 
second peak.  This suggests that an individual normalized light curve is related to the standard light curve $f(\lambda)$ as 
\begin{equation}\label{flexout}
\begin{split}
&l^J(t; \alpha, \beta, d, r)\\
& = \begin{cases}  d [f(\lambda(t)) - f_0], &  \lambda(t) \le \lambda_t \\
 d[f(\lambda_t) - f_0] + r[ f(\lambda(t)) - f(\lambda_t)], & \lambda(t) > \lambda_t
\end{cases}
\end{split}
\end{equation}
where  $f_0 = f(0) \equiv 0$, and the parameters $d$ and $r$ are positive constants of order one.  The decline parameter $d$ controls 
the depth of trough by scaling the decline from maximum light, ``vertically'' in the magnitude dimension.  A larger $d$ will produce a 
deeper trough and a faster decline rate (in magnitudes per day).  At the trough the magnitude is $J(T_{tr}) = J_0 + d 
[f(\lambda_t) - f_0]$ and rise parameter $r$ controls the rise in flux towards the second maximum relative to the trough magnitude.  A 
larger $r$ will produce a higher second peak and a faster rise rate.  This parameterization is constructed to preserve continuity in the 
light curve even as different phases are scaled in amplitude.  This quantitative parameterization of two constants $(\alpha, \beta)$ to 
control time scales and two constants $(d, r)$ to control amplitudes in two different regimes of the light curve is sufficient to describe the  
variation in $J$-band light curves.  This is a simple  transformation from the fiducial light curve to the realized light curves of 
 individual supernovae.

After the fiducial light curve $f(\lambda)$ is solved as a continuous function of feature time $\lambda$, one can easily measure any key 
features, such as the feature time of trough minimum ($\lambda_t$), the feature time of second peak ($\lambda_{p2}$), and the 
normalized magnitudes at these points, $f(\lambda_t)$ and $f(\lambda_{p2})$.   For any particular supernova's $J$-band light curve,  
we can measure these features using the solved parameters.  The chronological time of the trough is $T_{tr} = t(\lambda_t) = \alpha 
\lambda_t$, the trough-to-second-peak time is $T_2 - T_{tr} = \beta(\lambda_{p2} - \lambda_t)$, the depth of the trough is $J(T_{tr}) - 
J_0 = d \left[ f(\lambda_t) - f_0 \right]$ and the height of the second peak above the trough is $J(T_2) - J(T_{tr} ) =  r \left[ f(\lambda_{p2}) 
- f(\lambda_t) \right]$.  The parameters of the model can be directly related to observable features of light curves.

\subsubsection{Maximum Likelihood Construction of the FLIRT model}

The FLIRT model described above is completely specified by the fiducial normalized light curve function $f(\lambda)$.   We represent 
this function in the same way as the Maximum Likelihood Light Curve Template (\S \ref{MaxL}).  A set of knots $(\bm{p}, \bm{\tau})$ is 
defined on a regular grid, such that $f(\lambda) = \bm{s}(\lambda; \bm{\tau})\cdot\bm{p}$, where the vector $\bm{s}(\lambda; \bm{\tau})
$ is fully specified by natural cubic spline interpolation.  Once $\bm{p}$ is known, an individual supernova light curve can be fitted with 
the FLIRT model by means of nonlinear maximization of the likelihood to get point estimates of the light curve shape parameters $
\bm{\theta}^J_s =  (d^s, r^s, \alpha^s, \beta^s)$ and apparent magnitude $F_0^s$.  All that is now required is an estimate of the fiducial 
template $\bm{p}$.  The joint posterior over the supernova parameters and the fiducial template is proportional to the likelihood 
function
\begin{equation}
P( \{J_0^s, \bm{\theta}_s^J\}, \bm{p} | \,\{ \mathcal{D}^J_s \}) \propto \prod_{s=1}^{N_{\text{SN}}} N( \bm{m}^J_s | \, J_0^s + \bm{\tilde{S}}
(\bm{\theta}_s^J) \bm{p}, \, \bm{W}^F_s) 
\end{equation}
where the matrix $\bm{\tilde{S}}(\bm{\theta}_s^J)$ is  derived from the defining equations Eq. \ref{invtw}, Eq. \ref{flexout} and the 
interpolation method $\bm{s}(\lambda; \bm{\tau})$.  If $\{J_0^s, \bm{\theta}_s^J\}$ are estimated and fixed, then the conditional 
maximization of $\log P( \bm{p}\, | \,\{J_0^s, \bm{\theta}_s^J\}, \{ \mathcal{D}_s^J\})$ with respect to the template $\bm{p}$ subject to the 
constraint that $f(\lambda = 0) = p_j = 0$ is a linear problem.

It is straightforward to solve for the FLIRT model template iteratively.  We select a subset of SN light curves that are well sampled.  
First we pool the photometric data together and estimate the Maximum Likelihood Light Curve Template $\bm{\hat{p}}_0$ (\S 
\ref{MaxL}) as a first approximation.  Then we fit the FLIRT model using this template to each SN light curve by conditional 
maximization of $P( J_0^s, \bm{\theta}_s^J, | \, \bm{\hat{p}}_0  , \{\mathcal{D}^J\})$ to get estimates $\hat{J}_0^s, \bm{\hat{\theta}}_s^J$.  
Next we fix the SN parameters and update the fiducial template by constrained conditional maximization of $P( \bm{p}\, | \,\{\hat{J}_0^s, 
\bm{\hat{\theta}}_s^J\},  \{\mathcal{D}_s^J\})$ to get a new template $\bm{\hat{p}}_1$.  We iterate until the maximum likelihood template 
$\bm{\hat{p}}$ converges.   The template is rescaled so that the sample median values of the fitted SN 
shape parameters $(d, r, \alpha, \beta)$ are equal to one (so that the fiducial template model, which has shape parameters equal to 
one, reflects a typical light curve).

\begin{figure}[t]
\centering
\includegraphics[angle=0,scale=0.45]{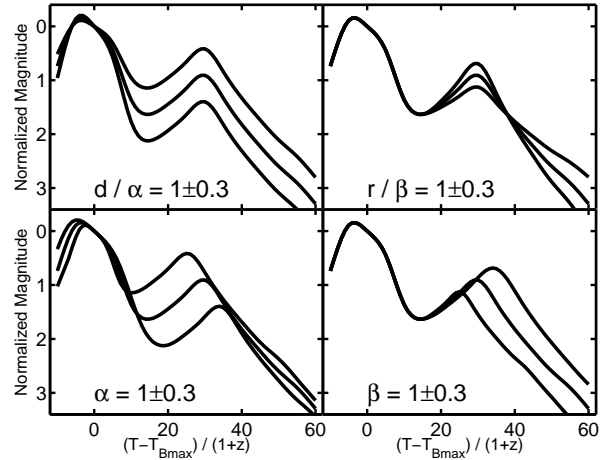}
\caption{\label{Jtemp} FLIRT model for $J$-band light curve shape variations.  In each panel, the fiducial FLIR template with 
parameters $(d/\alpha, r/\beta, \alpha, \beta) = (1,1,1,1)$ is shown as the middle curve along with models with one parameter varied 
while keeping the others fixed.  For example, the first panel depicts $(0.7, 1, 1,1)$ and $(1.3, 1, 1, 1)$.  The parameters correspond to 
the initial decline rate, the second rise rate, the time from peak to trough and the time from trough to second peak.}
\end{figure}

In practice, it is convenient to measure the relative decline rates $d / \alpha$ and the relative rise rates $r / \beta$ rather than the 
depths and heights directly.  The light curve shape parameters are then $\bm{\theta}^J = (d/\alpha, r/\beta, \alpha, \beta)$.  The 
maximum likelihood $J$-band FLIRT fiducial template $\bm{\hat{p}}_J$ is listed in Table 1, and depicted in Fig. \ref{Jtemp}, which  also 
shows the effects of varying each of the parameters.   The trough of the fiducial template is located at $(\lambda_t, f(\lambda_t) ) = 
(14.43 \text{ days}, 1.64 \text{ mag})$ and the second peak is  $(\lambda_{p2}, f(\lambda_{p2}) ) = (29.55 \text{ days}, 0.90 \text{ mag})
$.

In Fig. \ref{Jcomparison} we display the $J$-band FLIRT fiducial light curve, along with the $J$-band photometry for the $39$ SN listed 
in Table 2, shown in grey.  We have also transformed each SN light curve data set using the fitted light curve parameters in Table 2 to 
the same scales as the fiducial light curve by inverting Eq. \ref{flexout}.  The dramatic reduction of dispersion from 5 to 60 days shows that the 
FLIRT model successfully captures the shape variations in the $J$-band SN Ia light curves.  The double-peaked light curve structure is 
also seen in the $H$, $K$ and $I$ bands.  In the future, it may be worth exploring  FLIRT models  in those bands.

\begin{deluxetable}{rrrr}
\tabletypesize{\scriptsize}
\tablecaption{$JHK_s$ FLIRT and Max. Likelihood Templates\label{table1}}
\tablewidth{0pt}
\tablehead{ \colhead{$T-T_0$} & \colhead{$J-J_0$} & \colhead{$H-H_0$} & \colhead{$K_s-K_{s0}$}} 
\startdata
-10 &  0.74 &  0.34 &  0.48 \\ 
 -5 &  -0.11 &  -0.18 &  0.03 \\ 
 0 &  0.00 &  0.00 &  0.00 \\ 
 5 &  0.44 &  0.09 &  0.22 \\ 
 10 &  1.36 &  0.21 &  0.40 \\ 
 15 &  1.63 &  0.08 &  0.30 \\ 
 20 &  1.44 &  -0.13 &  0.18 \\ 
 25 &  1.15 &  \nodata & \nodata \\ 
 30 &  0.91 &  \nodata & \nodata \\ 
 35 &  1.33 &  \nodata & \nodata \\ 
 40 &  1.82 &  \nodata & \nodata \\ 
 45 &  2.23 &  \nodata & \nodata \\ 
 50 &  2.60 &  \nodata & \nodata \\ 
 55 &  2.91 &  \nodata & \nodata \\ 
 60 &  3.29 &  \nodata & \nodata \\ 
 \enddata
 \tablecomments{All templates are interpolated using natural cubic splines.}
\end{deluxetable}

\section{Application and Results}

\subsection{Nearby SN Ia NIR Light Curves}

A comprehensive data set of nearby SN Ia light curves in the near infrared was compiled by WV08, including observations of 21 recent 
SN with the Peters Automated InfraRed Imaging TELescope (PAIRITEL) taken by the CfA Supernova Group and observations of 23 SN 
from the literature \citep{jha99, hernandez00, krisciunas00, dipaola02, valentini03, krisciunas01, krisciunas03, krisciunas04a, 
krisciunas04b, krisciunas07,elias-rosa06,elias-rosa07, pastorello07a, stanishev07, pignata08}.  
Of these, three (SN 2005bl, SN 2005hk, and SN 2005ke) are omitted because they are fast-declining, peculiar SN with ``dromedary'' 
$H$-band light curves that have only one peak, whereas most $H$-band light curves are ``bactrian,'' having two peaks.  We use the remaining data set with 
two exceptions.  The very late $J$-band secondary maximum of SN 2002cv, and its extreme reddening and estimated optical extinction $
(A_V > 8)$ \citep{elias-rosa07} suggest this light curve is unusual, so we have omitted it from the analysis. We have also omitted the 
PAIRITEL observations of SN 2005eu, because we judged the $JHK_s$  image subtractions to be of poor quality.  The final, edited set of 
observations covers 39 SN Ia.  We have only used photometric measurements with signal-to-noise ratio $ > 3$.  Extensive studies of two SN in this set, SN 2005cf and SN 2006X, were presented by \citet{wangx08, wangx09}.

\begin{figure}[t]
\centering
\includegraphics[angle=0,scale=0.48]{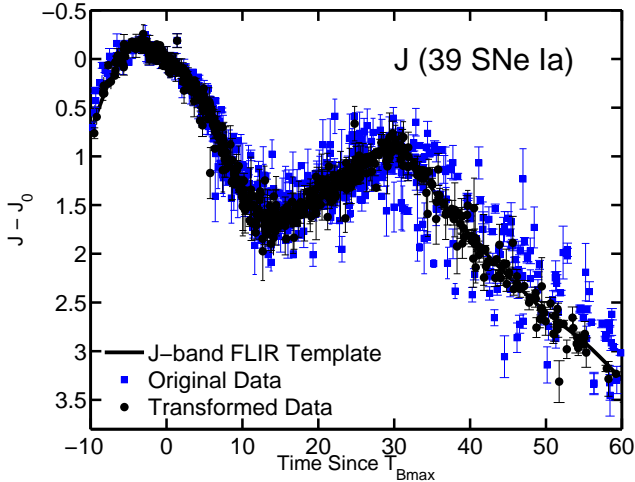}
\caption{\label{Jcomparison} The $J$-band light curve data (blue squares) exhibits significant shape variation after the initial decline.  
After fitting for these variations using the FLIRT model, the data was transformed to the fiducial frame and overplotted (black dots) with 
the fiducial FLIR template light curve (black line).  This demonstrates that the FLIRT model successfully captures these light curve 
shape variations.}
\end{figure}

To construct the $H$ and $K_s$-band templates (Fig. \ref{HKtemp}, Table 1) we used all the light curve observations from the data set.  
For the $J$-band light curves we selected a subset of 16 well-sampled light curves to generate the fiducial FLIRT template (Fig. 
\ref{Jtemp}, Table 1).  This subset consisted of   SN1998bu, 
SN 1999ee,
 SN 2001bt,
    SN 2001cn,
    SN 2001cz,
    SN 2002bo,
    SN 2005el,
    SN 2005eq,
    SN 2005na,
    SN 2006D,
   SN 2006N,
    SN 2006X,
    SN 2006ac,
    SN 2006ax,
    SN 2006le,
    and SN 2006lf.

All photometric data were K-corrected to the SN rest frame by linearly interpolating the tables of \citet{krisciunas04b}, and registered to 
a common phase by subtracting from the Julian Day the time of $B$-band maximum, $T_{Bmax}$, as determined by the MLCS2k2 fits 
to the optical light curves observed by the CfA Supernova Group \citep{hicken09a}.  The phases were corrected for time dilation using 
the heliocentric redshifts.  Recession velocities were corrected to the CMB+Virgo infall rest frame,  as described in WV08.  
Furthermore, a peculiar velocity uncertainty $\sigma_{pec} = 150 \text{ km s}^{-1}$ \citep{radburn-smith04} was assumed.  Luminosity 
distances were computed from the redshifts assuming an LCDM model with $\Omega_M = 0.27$, $\Omega_\Lambda = 0.73$ and a 
Hubble constant scale of $h = 0.72$ \citep{freedman01, spergel07}.  At the most distant end of the sample at $z\approx0.04$, the 
relative difference between the  luminosity distances in LCDM and in an Einstein-de Sitter universe is  2\%.

\subsection{$JHK_s$ Light Curve Model Specification}

The light-curve models we construct for the $JHK_s$ data set consist of maximum likelihood templates for $H$ and $K_s$ 
between -10 and 20 days (\S \ref{MaxL}, Fig. \ref{HKtemp}) and the $J$-band FLIRT model between -10 and 60 days (\S \ref{FLIRT}, 
Fig.  \ref{Jtemp}) with $\Delta \tau = 5$ days.  The $H$ and $K_s$ models have no light curve shape parameters, and the $J$-band 
FLIRT model has four:  $\bm{\theta}^H = \emptyset$, $\bm{\theta}^{K_s} = \emptyset$, and $\bm{\theta}^J = (d/\alpha, r/\beta, \alpha, 
\beta)$.  The multiband normalized light curve models as defined in Eq. \ref{gen_lc} are then fully specified by  
\begin{equation}
H(t) -H_0 = l_0^H(t) \equiv f_H(t) = \bm{S}(t; \tau) \cdot \bm{\hat{p}}^H, 
\end{equation}
\begin{equation}
K_s(t)  -K_{s0}= l_0^{K_s}(t) \equiv f_{K_s}(t) = \bm{S}(t; \tau) \cdot \bm{\hat{p}}^{K_s}, 
\end{equation}
\begin{equation}
J(t) -J_0 = \bm{l}_1^J(t; \bm{\theta}^J_{\text{NL}}) \cdot \bm{\theta}^J_{\text{L}} 
\end{equation}
where the $J$-band linear parameters are $\bm{\theta}^J_{\text{L}} = (d /\alpha, r /\beta)$, the nonlinear parameters are $
\bm{\theta}^J_{\text{NL}} = (\alpha, \beta)$, and the vector function $\bm{l}_1^J(t; \bm{\theta}^J_{\text{NL}})$ is determined by Eq. 
\ref{invtw} and Eq. \ref{flexout}.

In the notation of the hierarchical framework described in \S 2, the observable or apparent parameters  are $\bm{\phi}_s = 
(J_0, H_0, K_{s0}, d/\alpha, r/\beta, \alpha, \beta)$ for each for supernova $s$, and the intrinsic or absolute parameters are $
\bm{\psi}_s = (M_J, M_H, M_{K_s}, d/\alpha, r/\beta, \alpha, \beta)$ for each supernova $s$.  The population hyperparameters are $
\bm{\mu}_\psi = \mathbb{E}[ \bm{\psi}_s]$ and $\bm{\Sigma}_\psi = \text{Cov}[ \bm{\psi}_s, \bm{\psi}_s^T]$ with expectations with respect to 
the SN Ia NIR light curve population randomness.

Since dust extinction and reddening have small effect on the NIR light curves in our sample, we omit the full modeling of the multiband 
extinctions $\bm{A}_s$ and dust population characteristic $\tau_A$.   The most optically reddened SN in the sample (SN 1999cl, 
2006X, and SN 2003cg) are also at low redshifts, where the adopted velocity model gives them  little weight in the determinations of 
population means and covariances of the NIR absolute magnitudes.  Hence we set all $\bm{A}_s$ to zero and use the one-population 
model for  SN Ia NIR light curve randomness only.  In \S \ref{dustsection} we estimate the potential effect of dust on our posterior 
inferences.   In the near future, we will use the full two-population model with NIR and optical data for a simultaneous hierarchical 
modeling of SN Ia light curve shapes and dust extinction.

After plugging the specified $JHK_s$ light curve models and parameter dependence into the hierarchical framework of \S 2, we 
perform probabilistic inference  using the \textsc{BayeSN} algorithm of \S 2.4 to compute the joint posterior density over all individual 
parameters for the 39 SN and population hyperparameters.  There is a total of 347 parameters and hyperparameters in the statistical 
model.  Initial positions for the Markov chains were obtained by adding random noise to the MLE estimates of the SN parameters 
obtained in \S 3.  It is not necessary to specify initial guesses for the hyperparameters.   We set the scale of the inverse Wishart 
hyperprior, $\bm{\Lambda}_0 = \epsilon_0 \bm{I}$ by choosing a small value $\epsilon_0 = 10^{-4}$.   We found our inferences were insensitive 
to ten-fold changes in $\epsilon_0$.

The \textsc{BayeSN} MCMC algorithm was run for 5 independent chains with $2\times10^4$ samples each.  The Gelman-Rubin statistic was 
computed for all parameters:  the maximum value was 1.03 and 99\% had values less than 1.02, with the mean and median values 
less than 1.005.  Acceptable values of the G-R statistic are typically less than 1.10 \citep{gelman_bda}.  The first 2000 samples of each 
of the chains were then discarded as burn-in and the chains were concatenated for posterior analysis.  We found that our inferences 
were insensitive to the burn-in cutoff if it was greater than $\sim1000$ samples.

\subsection{Posterior Inferences}

The MCMC results produce samples from the global posterior density over all parameters and hyperparameters, Eq. 
\ref{globalposterior}.  We summarize the posterior density by examining marginal posterior densities over subsets of parameters.  
Inferences at the level of individual supernovae can be summarized by the probability density $P( \bm{\phi}_s, \mu_s | \, \mathcal{D}, 
\mathcal{Z})$ for each supernova $s$.  Inferences at the SN Ia NIR population level are summarized by $P( \bm{\mu}_\psi, 
\bm{\Sigma}_\psi | \, \mathcal{D}, \mathcal{Z})$.  This can be further broken down into the marginal densities over mean properties of 
absolute light curves $P(\bm{\mu}_\psi | \, \mathcal{D}, \mathcal{Z})$, the probability over covariances between multiband absolute 
magnitudes $P( \Sigma[ (M_J, M_H, M_{K_s}), (M_J, M_H, M_{K_s})] | \, \mathcal{D}, \mathcal{Z})$, marginal densities the covariances 
in light curve shape: $P(\Sigma(\bm{\theta}, \bm{\theta}))$ and marginal posterior densities over covariances between light curve 
shape and absolute magnitudes: $P( \Sigma[ (M_J, M_H, M_{K_s}), \bm{\theta}] | \, \mathcal{D}, \mathcal{Z})$.  These posterior 
densities are integrals over the global posterior density and can all be computed easily and directly from the MCMC chain.  We show 
example light curve data and model fits in Figs. \ref{sn2006X}, \ref{sn2006cp}, and \ref{sn2005cf}.

\subsubsection{SN Ia $JHK_s$ Light Curves}

The univariate marginal posterior median and standard deviations of the individual SN light curve parameters $\bm{\phi}_s  = (J_0, 
H_0, K_{s0}, d/\alpha, r/\beta, \alpha, \beta)$ for each of the 39 SN are listed in Table 2.  The light curve fits are excellent, especially in 
the $J$-band, where the  PAIRITEL photometry is the best.  The MCMC chains  quickly find the region of parameter space near  the 
peak of the posterior probability distribution, especially if the data tightly constrain the light curve fits.

The SN light curve data in the training set is not homogeneously well sampled.  Some SN, e.g. SN 2006X, have extremely good 
sampling in $JHK_s$ from before maximum to well past the secondary maximum.  Such well sampled, complete data sets constrain 
the observable light curve parameters very well.  Other SN are sparsely sampled or have incomplete coverage over the range of the 
model, for example SN 1999gp, SN 2007cq, SN 2005ao.  Some SN, for example SN 1998bu, are well sampled in the early part of the 
$J$-band light curve but the measurements stop before the secondary maximum.  Several SN in our sample (SN 1999ee, SN 2000bk, 
SN 2000ca, SN 2005iq, and SN 2007cq) have no $K_s$-band data.

\begin{figure}[b]
\centering
\includegraphics[angle=0,scale=0.5]{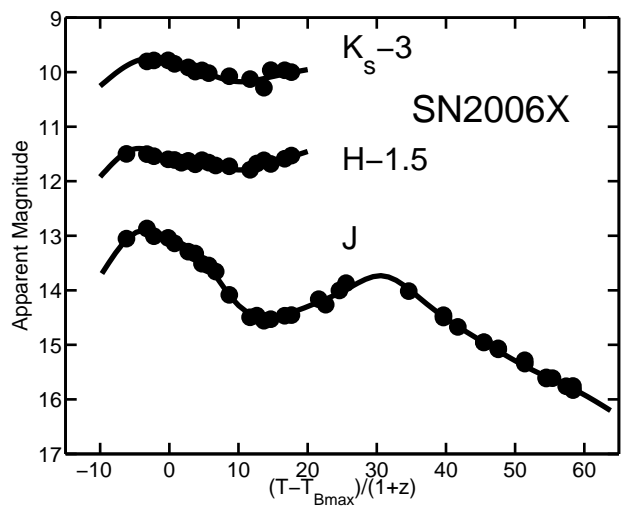}
\caption{\label{sn2006X} $JHK_s$ light curve data and model fits to SN 2006X.  This is a very well sampled light curve, and the fit to 
the light curve model (black curves) is excellent.  The $H$ and $K_s$ bands are fit to the maximum likelihood templates, and the $J$-
band is fit to the FLIRT model.  The data of this SN tightly constrain the light curve parameters.}
\end{figure}

\begin{figure}[h]
\centering
\includegraphics[angle=0,scale=0.5]{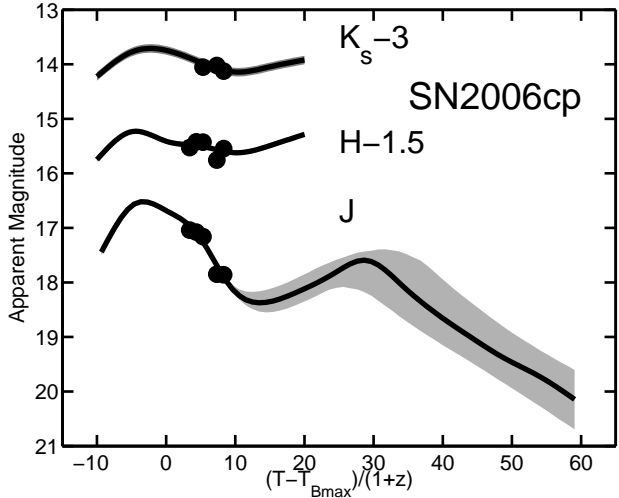}
\caption{\label{sn2006cp} $JHK_s$ light curve data and model fits (black curves) to SN 2006cp.  The light curve data is sparse and 
incomplete.  The \text{BayeSN} method estimates the $J$-band light curve where the data is missing using the information in the 
population distribution of the set of SN and its uncertainty.  For example, the correlation of the initial decline rate with the second rise 
rate provides some information.  Since the population of $J$-band light curves exhibits significant late-time shape variations, the late-
time model fit is very uncertain, as reflected by the grey error tube spanning the 16\% and 84\% quantiles of the posterior uncertainty in 
the light curve.}
\end{figure}

\begin{figure}[h]
\centering
\includegraphics[angle=0,scale=0.5]{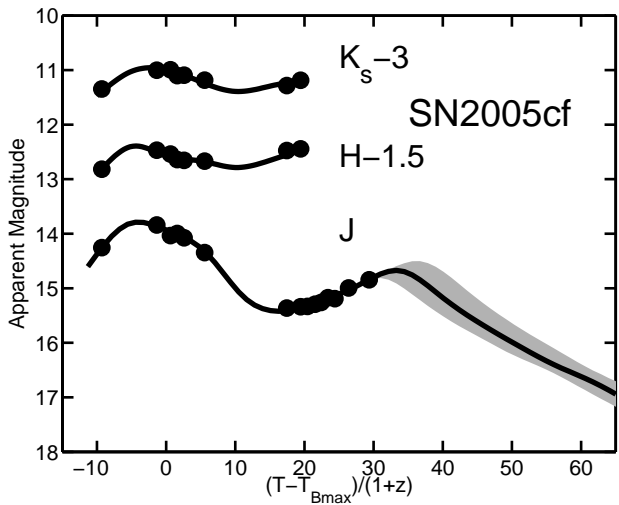}
\caption{\label{sn2005cf} $JHK_s$ light curve data and model fits (black curves) to SN 2005cf.  The light curve data is adequately 
sampled in the early part of the light curve up to second rise, but ends before reaching the second peak and decline.  The 
\text{BayeSN} method estimates the second maximum and late-time decline using a combination of the constraints imposed by the 
data and the population distribution of the training set.  For example, the final data point provides a lower bound for the time of the 
second maximum.  This makes the posterior distribution of the $\beta$ parameter non-gaussian.}
\end{figure}

In these cases the advantages of the Bayesian approach are clear.  Since we have defined a joint probability density over all data and 
parameters (both of which are considered random variables) in Eq. \ref{globalposterior}, we have a probability distribution over the 
parameters that are not well constrained by the individual SN data because of missing observations.  The Bayesian computation yields  
sensible numerical estimates of the poorly constrained parameters and their uncertainties using the joint probability over the observed 
parameters, the population distribution of individual SN parameters, and the uncertainty in the hyperparameters of the population 
distribution, all conditioned on the actual observed data and its uncertainty.  
For the very well-sampled, complete light curves, the posterior density over SN light curve parameters will be dominated by the 
information from its own light curve data.    An example of this is SN 2006X, shown in Fig. \ref{sn2006X} along with the light curve fits.
For sparse light curves, for example, SN 2006cp (Fig. \ref{sn2006cp}), some of the parameters will be informed by the population 
distribution constrained by the whole training set of SN.  In an intermediate case (e.g. SN 2005cf, Fig. \ref{sn2005cf},  an incomplete 
light curve), a balance between the existing data, observed parameters and the population distribution of the poorly constrained parameters is 
achieved, and an appropriate uncertainty is computed.  These computations are already handled automatically by our sampling of the 
global posterior probability density.  

\begin{deluxetable*}{lrrrrrrrrr} 
\tabletypesize{\scriptsize}
\tablecaption{ Posterior Summary of SN Ia $JHK_s$ Light Curve Parameters\label{table2}}
\tablewidth{0pt}
\tablehead{ \colhead{SN} & \colhead{$T_{Bmax}$}\tablenotemark{a} & \colhead{$J_0$}& \colhead{$H_0$} & \colhead{$K_{s0}$} & \colhead{$d/\alpha$} & \colhead{$r/\beta$} & \colhead{$\alpha$} & \colhead{$\beta$} & \colhead{Ref.} \tablenotemark{b}\\ 
 \colhead{} & \colhead{[MJD]} & \colhead{[mag]} & \colhead{[mag]} & \colhead{[mag]} & \colhead{} & \colhead{} & \colhead{} & \colhead{} & \colhead{}  }
\startdata
SN1998bu & 50952.40 & 11.75 (0.01) & 11.87 (0.01) & 11.58 (0.01) & 0.88 (0.03) & 0.94 (0.08) & 1.12 (0.02) & 1.02 (0.13) & J99,H00 \\ 
SN1999cl & 51342.20 & 12.94 (0.01) & 13.01 (0.01) & 12.66 (0.01) & 1.10 (0.03) & 1.12 (0.08) & 0.76 (0.02) & 1.05 (0.13) & K00 \\ 
SN1999cp & 51363.20 & 14.66 (0.02) & 14.93 (0.03) & 14.64 (0.07) & 0.85 (0.07) & 0.86 (0.08) & 1.15 (0.06) & 1.06 (0.26) & K00 \\ 
SN1999ee & 51469.30 & 14.96 (0.01) & 15.19 (0.01) & 14.94 (0.13) & 0.80 (0.01) & 0.85 (0.02) & 1.20 (0.01) & 1.08 (0.01) & K04b \\ 
SN1999ek & 51481.80 & 16.30 (0.01) & 16.42 (0.01) & 16.27 (0.09) & 0.97 (0.02) & 1.09 (0.07) & 1.00 (0.02) & 0.97 (0.26) & K04c \\ 
SN1999gp & 51550.10 & 16.95 (0.17) & 17.17 (0.12) & 16.75 (0.12) & 1.00 (0.18) & 0.84 (0.11) & 0.98 (0.17) & 1.10 (0.27) & K01 \\ 
SN2000E & 51577.20 & 13.58 (0.01) & 13.90 (0.02) & 13.49 (0.03) & 0.93 (0.02) & 0.88 (0.04) & 1.09 (0.01) & 1.16 (0.03) & V03 \\ 
SN2000bh & 51636.00 & 16.53 (0.03) & 16.88 (0.01) & 16.69 (0.02) & 1.03 (0.03) & 0.97 (0.02) & 1.06 (0.01) & 1.09 (0.01) & K04b \\ 
SN2000bk & 51647.00 & 17.15 (0.04) & 17.43 (0.01) & 17.52 (0.22) & 0.84 (0.04) & 1.23 (0.05) & 1.04 (0.02) & 0.58 (0.01) & K01 \\ 
SN2000ca & 51666.20 & 16.53 (0.01) & 16.78 (0.02) & 16.62 (0.20) & 0.83 (0.02) & 0.83 (0.09) & 1.14 (0.03) & 0.97 (0.25) & K04b \\ 
SN2000ce & 51667.30 & 15.98 (0.12) & 16.28 (0.02) & 15.87 (0.03) & 0.94 (0.11) & 0.89 (0.07) & 1.06 (0.04) & 1.05 (0.25) & K01 \\ 
SN2001ba & 52034.20 & 17.19 (0.01) & 17.51 (0.02) & 17.27 (0.02) & 1.03 (0.02) & 1.00 (0.04) & 1.06 (0.02) & 1.10 (0.03) & K04b \\ 
SN2001bt & 52062.90 & 15.55 (0.01) & 15.82 (0.02) & 15.51 (0.02) & 1.03 (0.02) & 0.96 (0.03) & 0.99 (0.02) & 0.96 (0.02) & K04c \\ 
SN2001cn & 52071.00 & 15.64 (0.03) & 15.91 (0.02) & 15.61 (0.05) & 0.96 (0.03) & 1.01 (0.03) & 1.09 (0.02) & 0.98 (0.02) & K04c \\ 
SN2001cz & 52103.40 & 15.53 (0.03) & 15.91 (0.05) & 15.63 (0.06) & 0.96 (0.04) & 0.89 (0.06) & 1.16 (0.04) & 1.08 (0.04) & K04c \\ 
SN2001el & 52182.10 & 13.03 (0.01) & 13.11 (0.01) & 12.86 (0.02) & 0.75 (0.02) & 0.93 (0.01) & 1.14 (0.01) & 0.88 (0.01) & K03 \\ 
SN2002bo & 52356.00 & 13.78 (0.02) & 14.08 (0.01) & 13.99 (0.02) & 0.90 (0.03) & 0.92 (0.03) & 1.02 (0.02) & 0.97 (0.02) & K04c \\ 
SN2002dj & 52450.60 & 14.68 (0.02) & 14.91 (0.01) & 14.64 (0.01) & 0.86 (0.02) & 0.88 (0.04) & 1.13 (0.02) & 0.92 (0.02) & P08 \\ 
SN2003cg & 52729.10 & 13.71 (0.04) & 13.92 (0.01) & 13.45 (0.01) & 1.04 (0.04) & 0.93 (0.04) & 0.95 (0.04) & 1.12 (0.03) & ER06 \\ 
SN2003du & 52765.90 & 14.29 (0.02) & 14.66 (0.02) & 14.35 (0.01) & 0.95 (0.02) & 0.96 (0.09) & 1.20 (0.02) & 1.04 (0.27) & St07 \\ 
SN2004S & 53038.70 & 14.82 (0.02) & 15.00 (0.01) & 14.71 (0.02) & 0.70 (0.01) & 0.85 (0.04) & 1.32 (0.03) & 0.83 (0.02) & K07 \\ 
SN2004eo & 53278.70 & 15.73 (0.04) & 15.97 (0.04) & 15.76 (0.08) & 1.09 (0.09) & 1.02 (0.04) & 0.87 (0.05) & 0.94 (0.05) & Pa07 \\ 
SN2005ao & 53442.00 & 17.98 (0.07) & 18.13 (0.01) & 18.34 (0.02) & 0.73 (0.17) & 1.06 (0.12) & 1.07 (0.14) & 0.88 (0.20) & WV08 \\ 
SN2005cf & 53533.60 & 13.93 (0.01) & 14.08 (0.01) & 13.99 (0.01) & 0.81 (0.02) & 0.91 (0.05) & 1.13 (0.02) & 1.12 (0.17) & WV08 \\ 
SN2005ch & 53536.00 & 17.03 (0.07) & 17.28 (0.03) & 17.07 (0.05) & 1.10 (0.08) & 1.05 (0.08) & 1.00 (0.04) & 0.99 (0.19) & WV08 \\ 
SN2005el & 53646.10 & 15.60 (0.01) & 15.82 (0.01) & 15.59 (0.01) & 1.13 (0.01) & 1.01 (0.02) & 0.87 (0.01) & 0.86 (0.01) & WV08 \\ 
SN2005eq & 53653.90 & 16.95 (0.01) & 17.34 (0.02) & 16.89 (0.03) & 0.76 (0.02) & 0.72 (0.03) & 1.19 (0.02) & 1.12 (0.02) & WV08 \\ 
SN2005iq & 53687.10 & 17.60 (0.05) & 17.79 (0.14) & 17.52 (0.20) & 1.06 (0.11) & 1.03 (0.10) & 0.92 (0.12) & 1.04 (0.26) & WV08 \\ 
SN2005na & 53740.50 & 16.66 (0.08) & 17.12 (0.07) & 16.85 (0.13) & 0.99 (0.12) & 0.89 (0.06) & 0.89 (0.05) & 1.21 (0.06) & WV08 \\ 
SN2006D & 53756.70 & 14.49 (0.01) & 14.70 (0.01) & 14.69 (0.01) & 1.05 (0.02) & 1.10 (0.04) & 0.98 (0.01) & 0.79 (0.02) & WV08 \\ 
SN2006N & 53760.60 & 15.69 (0.08) & 15.97 (0.07) & 15.81 (0.10) & 1.11 (0.09) & 1.10 (0.07) & 0.87 (0.04) & 0.93 (0.03) & WV08 \\ 
SN2006X & 53785.50 & 13.04 (0.02) & 13.08 (0.01) & 12.78 (0.01) & 0.91 (0.02) & 0.95 (0.01) & 0.97 (0.01) & 1.09 (0.01) & WV08 \\ 
SN2006ac & 53781.20 & 16.64 (0.07) & 16.90 (0.09) & 16.67 (0.09) & 0.94 (0.12) & 0.90 (0.08) & 0.88 (0.09) & 1.31 (0.19) & WV08 \\ 
SN2006ax & 53826.70 & 15.87 (0.01) & 16.37 (0.03) & 16.13 (0.03) & 1.08 (0.02) & 1.06 (0.06) & 1.12 (0.02) & 1.03 (0.02) & WV08 \\ 
SN2006cp & 53896.70 & 16.68 (0.04) & 16.91 (0.05) & 16.74 (0.08) & 1.10 (0.06) & 1.07 (0.07) & 0.94 (0.08) & 1.00 (0.26) & WV08 \\ 
SN2006gr & 54014.00 & 17.94 (0.10) & 18.00 (0.14) & 17.63 (0.18) & 0.91 (0.09) & 1.02 (0.10) & 1.03 (0.12) & 1.01 (0.27) & WV08 \\ 
SN2006le & 54048.00 & 16.35 (0.03) & 16.64 (0.02) & 16.22 (0.05) & 0.90 (0.05) & 0.96 (0.05) & 1.16 (0.04) & 1.14 (0.04) & WV08 \\ 
SN2006lf & 54044.80 & 15.78 (0.03) & 15.84 (0.04) & 15.56 (0.05) & 1.12 (0.07) & 1.12 (0.08) & 0.85 (0.05) & 0.79 (0.05) & WV08 \\ 
SN2007cq & 54280.00 & 16.50 (0.03) & 17.01 (0.24) & 16.80 (0.28) & 0.96 (0.21) & 0.89 (0.15) & 0.97 (0.30) & 1.10 (0.29) & WV08 \\ 
\enddata

\tablenotetext{a}{Time of maximum light in $B$: Julian Date - 2,400,000. $J_0, H_0, K_{s0}$ are $J$, $H$, and $K_{s}$ at $T_{Bmax}$.   }

\tablenotetext{b}{Reference codes: WV08 \citet[PAIRITEL photometry; ][]{wood-vasey08}; J99: \citet{jha99};
H00: \citet{hernandez00}; K00: \citet{krisciunas00}; K01: \citet{krisciunas01}; 
V03: \citet{valentini03}; K03: \citet{krisciunas03}; K04b: \citet{krisciunas04b};
K04c: \citet{krisciunas04c}; K07: \citet{krisciunas07}; ER06: \citet{elias-rosa06}; Pa07: \citet{pastorello07a}; St07: \citet{stanishev07}; P08: \citet{pignata08}.}

 \end{deluxetable*}

\begin{deluxetable*}{lrrrrrrr}
\tabletypesize{\scriptsize}
\tablecaption{Summary of Posterior Inference: Population Hyperparameters\label{table3}}
\tablewidth{0pt}
\tablehead{ \colhead{$\psi^i$} & \colhead{$M_J$} & \colhead{$M_H$} & \colhead{$M_{Ks}$} & \colhead{$d/\alpha$} & \colhead{$r/\beta$} & \colhead{$\alpha$} & \colhead{$\beta$} }
\startdata
$\mu(\cdot)$ &  -18.25 (0.03) & -18.01 (0.03) & -18.25 (0.04) & 0.95 (0.03) & 0.97 (0.03) & 1.04 (0.03) & 1.01 (0.04) \\ 
 $\sigma(\cdot)$ &  0.17 (0.03) & 0.11 (0.03) & 0.19 (0.04) & 0.15 (0.03) & 0.13 (0.03) & 0.15 (0.02) & 0.22 (0.04) \\ 
 \tableline
\tableline
$\rho(M_J,\cdot)$ &  1.00  & 0.73 (0.03) & 0.41 (0.09) & -0.14 (0.29) & 0.52 (0.03) & -0.07 (0.46) & -0.28 (0.18) \\ 
$\rho(M_H,\cdot)$ &  0.73 (0.03) & 1.00  & 0.53 (0.04) & -0.03 (0.47) & 0.59 (0.03) & 0.24 (0.20) & -0.21 (0.35) \\ 
$\rho(M_{Ks},\cdot)$ &  0.41 (0.09) & 0.53 (0.04) & 1.00  & -0.16 (0.34) & 0.76 (0.01) & 0.07 (0.39) & -0.48 (0.11) \\ 
$\rho(d/\alpha,\cdot)$ &  -0.14 (0.29) & -0.03 (0.47) & -0.16 (0.34) & 1.00  & 0.55 (0.02) & -0.77 (0.00) & 0.07 (0.41) \\ 
$\rho(r/\beta,\cdot)$ &  0.52 (0.03) & 0.59 (0.03) & 0.76 (0.01) & 0.55 (0.02) & 1.00  & -0.50 (0.02) & -0.43 (0.07) \\ 
$\rho(\alpha,\cdot)$ &  -0.07 (0.46) & 0.24 (0.20) & 0.07 (0.39) & -0.77 (0.00) & -0.50 (0.02) & 1.00  & -0.04 (0.47) \\ 
$\rho(\beta,\cdot)$ &  -0.28 (0.18) & -0.21 (0.35) & -0.48 (0.11) & 0.07 (0.41) & -0.43 (0.07) & -0.04 (0.47) & 1.00  \\ 
\enddata

\tablecomments{ (top) Population means and variances of the absolute parameters.  Values in the parentheses are the standard deviations of the marginal posterior density in each parameter.
The estimates of the $\sigma(\cdot)$ are modal values.  (bottom)  Population correlation matrix for the absolute parameters.  Estimates of the correlations $\rho(\cdot, \cdot)$ are the modal values.  The parentheses contain
the tail probabilities as described in Eq. \ref{ptail}.}

\end{deluxetable*}

\subsubsection{NIR Absolute Magnitudes}

A summary of posterior inferences of the SN Ia NIR light curve population hyperparameters is presented in Table 3.  The univariate 
expectation of the means $\bm{\mu}_\psi$ is shown along with the standard deviations of the univariate marginal densities.  We list the 
modal values of the square root of the variances $\sigma^2_\psi$, which are the diagonal values of the covariance matrix $
\bm{\Sigma}_\psi$, and the standard deviations of their univariate marginal posterior probability densities.  We also list the modal 
values of the correlations $\rho(\cdot, \cdot)$ obtained from the off-diagonal terms of the covariance $\bm{\Sigma}_\psi$ after factoring 
out the variances.  The marginal modes are estimated from the histogram of MCMC samples in each quantity. In addition we list the tail 
probabilities of each correlation coefficient, defined as
\begin{equation}\label{ptail}
p_{\text{tail}} = \begin{cases} P(\rho < 0), & \text{if mode}(\rho) > 0, \\
P(\rho > 0), & \text{if mode}(\rho) < 0.
\end{cases}
\end{equation}
The smaller the tail probability, the greater the evidence that the correlation is different from zero, either positively or negatively.  The 
probability densities of correlation coefficients have support between -1 and 1 and are typically asymmetric. The probability densities 
of variance parameters are also non-gaussian, since they are forced to be positive, and have fat tails towards higher variance.  This 
captures the intuition that for a finite sample with fixed scatter, it is more difficult to discount the hypothesis that it arose from a high-
variance distribution rather than a low variance one.

\begin{figure}[t]
\centering
\includegraphics[angle=0,scale=0.45]{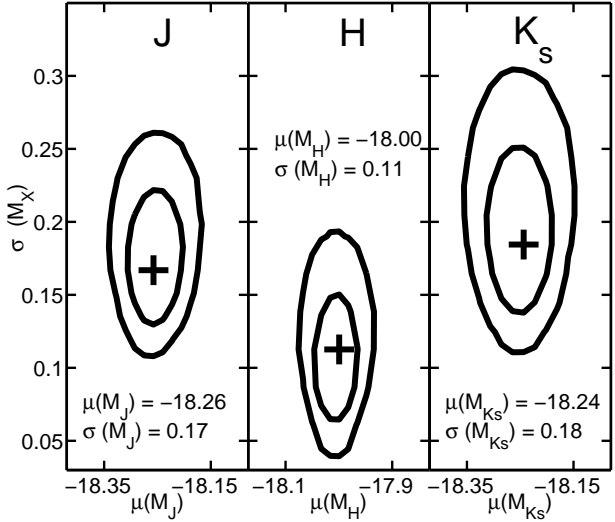}
\caption{\label{mvJHK}Joint posterior probability densities in the population mean $\mu$ and population variance $\sigma^2$ of the 
peak absolute magnitudes in each NIR band.  The crosses and the numbers in each panel
denote  the mode of the bivariate probability density.  The contours contain the 68\% and 95\% highest posterior density regions.  
These estimates were obtained directly from the \textsc{BayeSN} MCMC chain of the trained statistical model.    The univariate 
marginal estimates of the population variances are:  $\sigma(M_J) = 0.17 \pm 0.03$, $\sigma(M_H) = 0.11 \pm 0.03$, and $
\sigma(M_{Ks}) = 0.19 \pm 0.04$. }
\end{figure}

The population mean absolute magnitudes are $\mu(M_J) = -18.25 \pm 0.03$, $\mu(M_H) = -18.01 \pm 0.03$, and $\mu(M_{Ks}) = 
-18.25 \pm 0.04$ mag (on the scale of $h = 0.72$), and the population standard deviations are $\sigma(M_J)  = 0.17 \pm 0.03$, $
\sigma(M_H) = 0.11 \pm 0.03$, and $\sigma(M_{Ks}) = 0.19 \pm 0.04$ mag.  In Figure \ref{mvJHK}, we show the bivariate joint 
posterior density of the mean and variance for the absolute magnitude in each band, and the bivariate modal values.  The skews in the 
posterior densities for the variances are visible.  The absolute magnitude in the $H$-band clearly has much less intrinsic dispersion 
than in the $J$- and $K$-band and is the best constrained.  We have used bivaraite kernel density estimation with the MCMC samples 
to compute the 68\% and 95\% highest posterior density contours and the mode, shown in the figure.

\begin{figure}[t]
\centering
\includegraphics[angle=0,scale=0.45]{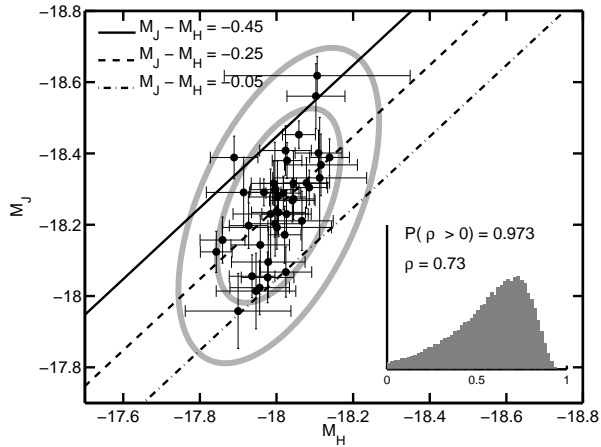}
\caption{\label{MJvsMH} Strong correlation $\rho$ between $J$- and $H$- band peak luminosities.  The grey ellipses contain 95\% 
and 68\% of the bivariate population probability distribution using the modal values of the population covariance.  The straight lines 
indicate sets of constant $J-H$ color.  There appears to be a trend that bluer $J-H$ objects are also intrinsically brighter.  The marginal 
posterior probability density of the correlation coefficient obtained via MCMC is shown in the inset.  The mode is $\rho = 0.73$, and 
$P(\rho > 0) = 0.97$ is obtained by numerical integration of the marginal density.}
\end{figure}

 Figure \ref{MJvsMH} shows the marginal posterior estimates of the individual SN $H$ and $J$ absolute magnitudes (obtained from, 
e.g. $P(J_0^s - \mu_s | \, \mathcal{D}, \mathcal{Z})$) plotted with contours representing the 68\% and 95\% probability contours of the 
bivariate population density  $P( M_J, M_H |\, \bm{\mu}_\psi, \bm{\Sigma}_\psi)$ estimated using the modal values of the covariance 
matrix $\bm{\Sigma}_\psi$ (Table 3).  We also show the marginal posterior density of the correlation coefficient for the pair of absolute 
magnitudes.

We see that the absolute magnitudes in $J$ and $H$ are highly correlated ($\rho \approx 0.73$) with strong evidence for positive 
correlation $(P(\rho > 0) > 0.97)$.  The data also suggest that intrinsically brighter SN are typically bluer in $J-H$ color.  Interestingly, 
this parallels the ``brighter-bluer'' relation seen in optical light curves \citep[e.g.][]{guy05,jha06}.  There is also evidence for positive 
correlations between $M_J$, $M_{Ks}$ and $M_H$, $M_{Ks}$, although the modal correlations are weaker ($\rho \approx 0.4 - 0.5$, 
Table 3).

\subsubsection{Statistical Structure of $J$-band Light Curve Shapes}

We examine the statistical relationships between the different features of the $J$-band light curve.  We focus only on those correlations 
which are significantly different from zero, as measured by the tail probabilities of the posterior distribution (Table 3).

The peak-to-trough initial decline rate, as measured by $d / \alpha$, is moderately correlated ($\rho \approx 0.55$) with the trough-to-
second peak rise rate, as measured by  $r / \beta$.  The posterior probability of a positive correlation is 98\%.  This trend indicates  
faster pre-trough declines lead to faster post-trough rises.  There is a strong correlation ($\rho \approx -0.77$) in the early light curve 
between the initial decline rate and the time to trough ($\alpha$), demonstrating that slower declining light curves have later troughs.  
There is a moderate ($\rho \approx -0.43$) correlation between the post-trough rise rate and the trough-to-second peak time ($\beta$), 
suggesting that in the post-trough phase, slower risers have later secondary maxima.  The posterior probability of a negative 
correlation is 93\%.  As shown in Table 3, there is no correlation between the early time scale ($\alpha$) and the late, post-trough time 
scale ($\beta$).

\subsubsection{Statistical Correlations between NIR Absolute Magnitudes and Light Curve Shape}

Statistical correlations between peak SN Ia absolute magnitudes and light curve shape are of paramount importance to cosmological 
studies, because they relate the intrinsic luminosity,  a hidden, intrinsic parameter, to a distance-independent observable 
measure.   Relations between optical light curve shape and optical luminosity have been leveraged to improve the utility of SN Ia 
as standard candles and distance indicators.  We present the first quantitative search and measurement of correlations between near 
infrared absolute magnitudes and light curve shape as measured from the $J$-band light curves.  Again, we only highlight correlations 
having the highest posterior probability of being non-zero as measured from the tail probability.

\begin{figure}[t]
\centering
\includegraphics[angle=0,scale=0.45]{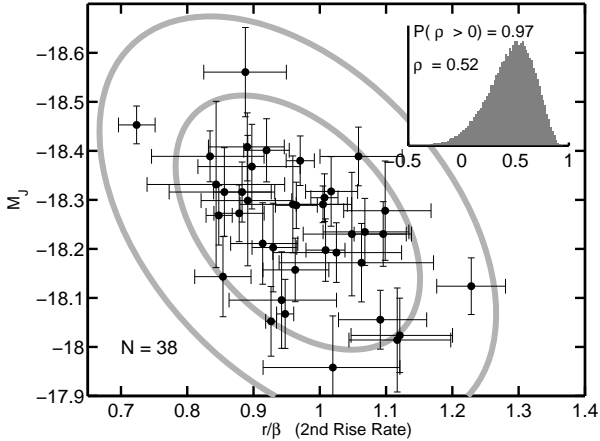}
\caption{\label{MJvsRise}Evidence for moderate correlation between the $J$-band second rise rate and the $J$-band peak luminosity.  
The grey ellipses contain 95\% and 68\% of the bivariate population probability distribution using the modal values of the population 
covariance. The inset shows the marginal posterior probability density of the correlation coefficient obtained via MCMC, along with the 
mode and probability of positive correlation with the absolute magnitude.   }
\end{figure}

 Figure \ref{MJvsRise} show the $M_J$ and $r / \beta$ estimates for individual SN together with 68\% and 95\% probability 
contours of the population density  $P( M_J, r/\beta |\, \bm{\mu}_\psi, \bm{\Sigma}_\psi)$ using the expected posterior estimate of the 
population mean and the modal covariance matrix.  The most likely correlation is moderate ($\rho \approx 0.52$).  The evidence for a 
positive correlation is fairly strong: $P(\rho > 0) = 97\%$.  This demonstrates that brighter SN Ia $J$-band light curves are likely to rise 
more slowly to the second maximum.

\begin{figure}[h]
\centering
\includegraphics[angle=0,scale=0.45]{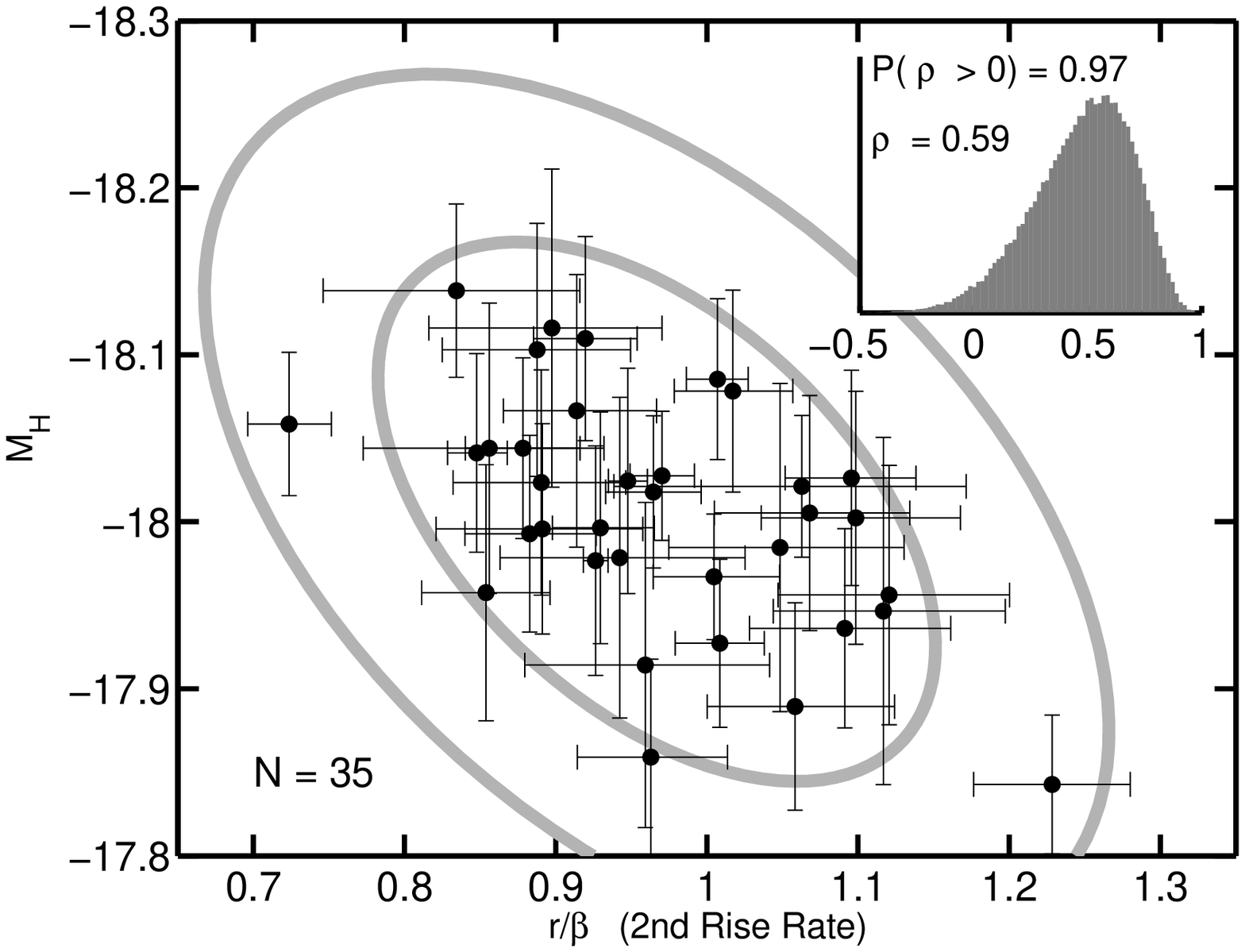}
\caption{\label{MHvsRise} Moderate correlation between the $J$-band second rise rate and the $H$-band peak luminosity.  The grey 
ellipses contain 95\% and 68\% of the bivariate population probability distribution using the modal values of the population covariance. 
The inset shows the marginal posterior probability density of the correlation coefficient obtained via MCMC, along with the mode and 
probability of positive correlation with the absolute magnitude. }
\end{figure}

\begin{figure}[h]
\centering
\includegraphics[angle=0,scale=0.45]{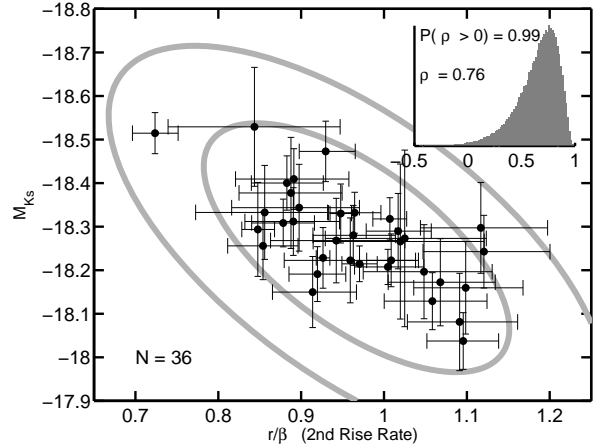}
\caption{\label{MKvsRise} Evidence for correlation between the $J$-band second rise rate and the $K_s$-band peak luminosity.  The 
grey ellipses contain 95\% and 68\% of the bivariate population probability distribution using the modal values of the population 
covariance. The inset shows the marginal posterior probability density of the correlation coefficient obtained via MCMC, along with the 
mode and probability of positive correlation with the absolute magnitude.  }
\end{figure}

Figure \ref{MHvsRise}, shows  a moderate correlation ($\rho \approx 0.59$) of $J$-band rise rate with the $H$-band luminosity.  
There is good evidence for a positive correlation ($P(\rho > 0) = 0.97$).   Figure \ref{MKvsRise} shows that the $K_s$-band luminosity 
has a fairly strong correlation with the $J$-band rise rate ($\rho \approx 0.76$) with strong evidence ($P(\rho > 0) = 99\%$) for a 
positive relation.   In these bivariate plots we have only shown individual SN with posterior uncertainty smaller than the population 
width in each parameter.  The fully Bayesian calculation properly accounts for the uncertainties in the parameters of individual SN 
when  determining the posterior density of the population correlation.

Taken together these correlations suggest that SN Ia light curves brighter in the NIR at peak have slower rates of evolution at later 
times, as measured from the $J$-band light curve.   A larger sample of SN Ia light curves in the NIR is needed to confirm these 
correlations.  The best measured light curves tend to be at lower redshifts, where peculiar velocity uncertainties make the absolute 
magnitudes highly uncertain.  Supernovae  farther out in the Hubble flow have better determined absolute magnitudes but are likely to 
have poorer quality measurements of the whole light curve evolution.  Continued monitoring of local SN in the NIR over a wide range 
of distances and redshifts will help to solidify our estimates of these correlations (by narrowing their posterior probability densities and 
providing better estimates of the modal correlation coefficients).

Using the trained statistical model, we can estimate the expected precision of distance prediction for different types of light curve 
observations represented by subsets of the observable light curve parameters in $\bm{\phi}_s$.   Suppose the observable vector $
\bm{\tilde{\phi}}_s$ of a new hypothetical SN can be partitioned into the observed parameters and the unobserved parameters $
\bm{\tilde{\phi}}_s = (\bm{\tilde{\phi}}_s^o, \bm{\tilde{\phi}}_s^u)$.   We have computed the  variance of the predictive $\tilde{\mu}_s$ 
conditional on $\bm{\tilde{\phi}}_s^o$, and marginalizing over  $\bm{\tilde{\phi}}_s^u$ and the posterior uncertainty of the 
hyperparameters $\bm{\mu}_\psi$, $\bm{\Sigma}_\psi$ for various partitions of $\bm{\tilde{\phi}}_s$.    We find that the statistical model 
implies the following properties:
(1) If one only observes the light curve around $T_{Bmax}$, then the single most valuable measurement is the $H$-band apparent 
magnitude, providing a distance modulus precision of $\sim 0.14$ mag; the $J$- and $K_s$-bands do not add much more statistical 
power. (2) If one monitors the $J$-band light curve at late times to measure the second rise to the secondary maximum, the moderate 
correlation of the rise rate with absolute magnitudes can be used to decrease the uncertainty towards $\sim 0.1$ mag.  In the next 
section, we will test the predictive performance of the model using the NIR SN Ia sample.

\subsection{The Hubble Diagram of $JHK_s$ SN Ia Light Curves}

\subsubsection{Hubble Residuals and Training Error}

We list in Table 4 the redshifts and several estimates of the distance moduli to the SN Ia in our training set.  We list the redshift-based 
LCDM Hubble flow distance and its uncertainty, described by Eq. \ref{gaussianmuprior}, on the $H_o = 72 \text{ km s}^{-1} 
\text{ Mpc}^{-1}$ scale.  This describes the factor $P(\mu_s | z_s)$, conditioning on the redshift only, and incorporating the assumed 
$\sigma_\text{pec} = 150 \text{ km s}^{-1}$ peculiar velocity uncertainty and the redshift measurement error.  As a product of the 
Bayesian treatment, we obtain the posterior estimate of the distance modulus, combining the redshift information with the statistical light 
curve model and conditioning on the entire dataset to generate an ``information update.''  This is expressed as the 
marginal posterior probability $P( \mu_s | \, \mathcal{D}, \mathcal{Z})$.  The mean and standard deviation of this probability density are 
listed as $\mu_{\text{post}}$ and $\sigma_\text{post}$ for each supernova.

\begin{deluxetable*}{lrrrrrrrrrrr}
\tabletypesize{\scriptsize}
\tablecaption{SN Ia NIR Distance Modulus Estimates\label{table4}}
\tablewidth{0pt}
\tablehead{ \colhead{SN} & \colhead{$cz$}\tablenotemark{a}  & \colhead{$cz_{\text{err}}$} & \colhead{$\mu_{LCDM}|z$} & \colhead{$\sigma_{\mu}|z$} & \colhead{$\mu_{\text{post}}$}\tablenotemark{b} & \colhead{$\sigma_{\text{post}}$} & \colhead{$\mu_{\text{resub}}$} & \colhead{$\sigma_{\text{resub}}$} & \colhead{$\bar{\mu}_{\text{pred}}$} & \colhead{$s_{\text{pred}}$}& \colhead{$\sigma_{\text{pred}}$}\\
\colhead{} 
  & \colhead{[$\text{km s}^{-1}$]}    & \colhead{[$\text{km s}^{-1}$]}               & \colhead{[mag]} 
  &  \colhead{[mag]} & \colhead{[mag]}  & \colhead{[mag]}              & \colhead{[mag]} 
  & \colhead{[mag]}   & \colhead{[mag]}       & \colhead{[mag]}                   & \colhead{[mag]} }
\startdata
SN1998bu & 709 & 20 & 30.00 & 0.46 & 29.84 & 0.10 & 29.85 & 0.09 & 29.84 & 0.09 & 0.12 \\ 
SN1999cl & 957 & 86 & 30.62 & 0.39 & 30.95 & 0.10 & 30.99 & 0.08 & 31.00 & 0.12 & 0.14 \\ 
SN1999cp & 2909 & 14 & 33.05 & 0.11 & 32.98 & 0.08 & 32.94 & 0.14 & 32.94 & 0.04 & 0.15 \\ 
SN1999ee & 3296 & 15 & 33.32 & 0.10 & 33.23 & 0.06 & 33.20 & 0.05 & 33.16 & 0.07 & 0.09 \\ 
SN1999ek & 5191 & 10 & 34.32 & 0.06 & 34.36 & 0.06 & 34.45 & 0.14 & 34.47 & 0.06 & 0.15 \\ 
SN1999gp & 8113 & 18 & 35.30 & 0.04 & 35.28 & 0.04 & 35.10 & 0.20 & 35.05 & 0.06 & 0.20 \\ 
SN2000E & 1803 & 19 & 32.00 & 0.18 & 31.89 & 0.06 & 31.89 & 0.07 & 31.86 & 0.09 & 0.11 \\ 
SN2000bh & 6765 & 21 & 34.90 & 0.05 & 34.91 & 0.04 & 34.90 & 0.04 & 34.88 & 0.04 & 0.06 \\ 
SN2000bk & 7976 & 20 & 35.27 & 0.04 & 35.27 & 0.04 & 35.34 & 0.08 & 35.57 & 0.13 & 0.15 \\ 
SN2000ca & 6989 & 62 & 34.97 & 0.05 & 34.92 & 0.05 & 34.78 & 0.10 & 34.75 & 0.05 & 0.11 \\ 
SN2000ce & 5097 & 15 & 34.28 & 0.06 & 34.28 & 0.06 & 34.28 & 0.12 & 34.24 & 0.09 & 0.14 \\ 
SN2001ba & 8718 & 22 & 35.46 & 0.04 & 35.48 & 0.03 & 35.53 & 0.08 & 35.51 & 0.06 & 0.10 \\ 
SN2001bt & 4220 & 13 & 33.87 & 0.08 & 33.84 & 0.05 & 33.83 & 0.05 & 33.82 & 0.06 & 0.08 \\ 
SN2001cn & 4454 & 250 & 33.97 & 0.14 & 33.84 & 0.05 & 33.83 & 0.06 & 33.86 & 0.05 & 0.08 \\ 
SN2001cz & 4506 & 20 & 34.00 & 0.07 & 33.94 & 0.06 & 33.88 & 0.12 & 33.85 & 0.04 & 0.12 \\ 
SN2001el & 978 & 10 & 30.70 & 0.33 & 31.08 & 0.07 & 31.10 & 0.03 & 31.17 & 0.08 & 0.08 \\ 
SN2002bo & 1696 & 20 & 31.88 & 0.19 & 32.19 & 0.06 & 32.19 & 0.05 & 32.21 & 0.11 & 0.12 \\ 
SN2002dj & 2880 & 22 & 33.03 & 0.11 & 32.95 & 0.05 & 32.94 & 0.05 & 32.92 & 0.07 & 0.09 \\ 
SN2003cg & 1340 & 24 & 31.37 & 0.25 & 31.92 & 0.07 & 31.97 & 0.06 & 31.97 & 0.18 & 0.19 \\ 
SN2003du & 2206 & 14 & 32.44 & 0.15 & 32.58 & 0.10 & 32.61 & 0.17 & 32.60 & 0.08 & 0.19 \\ 
SN2004S & 2607 & 16 & 32.81 & 0.13 & 32.96 & 0.08 & 33.01 & 0.06 & 33.07 & 0.12 & 0.14 \\ 
SN2004eo & 4859 & 17 & 34.17 & 0.07 & 34.05 & 0.05 & 33.93 & 0.10 & 33.92 & 0.05 & 0.11 \\ 
SN2005ao & 11828 & 126 & 36.14 & 0.04 & 36.15 & 0.04 & 36.20 & 0.10 & 36.40 & 0.20 & 0.23 \\ 
SN2005cf & 2018 & 11 & 32.24 & 0.16 & 32.14 & 0.08 & 32.13 & 0.06 & 32.10 & 0.09 & 0.11 \\ 
SN2005ch & 8094 & 1499 & 35.30 & 0.40 & 35.26 & 0.10 & 35.27 & 0.11 & 35.25 & 0.06 & 0.12 \\ 
SN2005el & 4349 & 8 & 33.93 & 0.08 & 33.91 & 0.05 & 33.90 & 0.03 & 33.85 & 0.08 & 0.08 \\ 
SN2005eq & 8535 & 25 & 35.41 & 0.04 & 35.40 & 0.04 & 35.39 & 0.05 & 35.26 & 0.15 & 0.16 \\ 
SN2005iq & 10102 & 40 & 35.79 & 0.03 & 35.79 & 0.03 & 35.86 & 0.17 & 35.81 & 0.06 & 0.18 \\ 
SN2005na & 7826 & 26 & 35.23 & 0.04 & 35.22 & 0.04 & 35.21 & 0.14 & 35.16 & 0.10 & 0.17 \\ 
SN2006D & 2560 & 18 & 32.76 & 0.13 & 32.73 & 0.06 & 32.72 & 0.06 & 32.76 & 0.12 & 0.14 \\ 
SN2006N & 4468 & 27 & 33.99 & 0.07 & 33.97 & 0.06 & 33.98 & 0.15 & 33.95 & 0.06 & 0.16 \\ 
SN2006X & 1091 & 20 & 30.88 & 0.30 & 31.10 & 0.07 & 31.12 & 0.03 & 31.12 & 0.09 & 0.10 \\ 
SN2006ac & 7123 & 17 & 35.01 & 0.05 & 35.01 & 0.05 & 34.98 & 0.14 & 34.99 & 0.09 & 0.17 \\ 
SN2006ax & 4955 & 20 & 34.21 & 0.07 & 34.26 & 0.06 & 34.31 & 0.09 & 34.38 & 0.06 & 0.11 \\ 
SN2006cp & 6816 & 14 & 34.92 & 0.05 & 34.92 & 0.05 & 34.92 & 0.14 & 34.90 & 0.04 & 0.14 \\ 
SN2006gr & 10547 & 22 & 35.89 & 0.03 & 35.90 & 0.03 & 36.00 & 0.19 & 36.02 & 0.09 & 0.21 \\ 
SN2006le & 5403 & 12 & 34.40 & 0.06 & 34.50 & 0.06 & 34.63 & 0.08 & 34.69 & 0.04 & 0.09 \\ 
SN2006lf & 4048 & 10 & 33.77 & 0.08 & 33.80 & 0.07 & 33.83 & 0.11 & 33.85 & 0.17 & 0.20 \\ 
SN2007cq & 7501 & 50 & 35.13 & 0.05 & 35.11 & 0.05 & 35.01 & 0.19 & 34.91 & 0.03 & 0.19 \\ 
\enddata

\tablenotetext{a}{Corrected to the CMB+Virgo frame.}
\tablenotetext{b}{$\mu_{\text{post}}$ and $\sigma_\text{post}$ are the mean and standard deviation of the trained posterior density in the distance modulus.  $\mu_\text{resub}$ and $\sigma_\text{resub}$ are the mean and standard deviation of the resubstituted predictive posterior of the distance modulus.  $\bar{\mu}_\text{pred}$ and $s_\text{pred}$ are the average and scatter over multiple bootstrapped training sets of the expected predictive distance moduli.  $\sigma_\text{pred}^2$ is the quadrature sum of predictive uncertainty and the scatter over bootstrap predictions.}

\end{deluxetable*}

The typical measure of the quality of a model for SN Ia standard candles is the average residual in the Hubble diagram.  First, the 
redshift-based distance moduli and photometric light curves are used to ``train'' a statistical model, by determining the mean absolute 
magnitude and variance, and perhaps relationships between absolute magnitude and light curve shape.  Once these parameters of 
the statistical model are determined from the training set data, the photometric light curves are fed into the model \emph{without} the 
redshift-based distances to ``predict'' standard candle distances using the model.  These new distances are compared to the Hubble 
flow distances to calculate the average residual error.

This measure of ``training error'' may be called the \emph{resubstitution error} because it involves using the redshifts and light curve 
data of the training set to train the model parameters, and then the resubstituting the light curve data back into the model to produce 
distance ``predictions''  \emph{as if} the light curve data were new.

In our fully Bayesian formulation, the process of ``training'' corresponds to computing the (non-gaussian) posterior density over the 
hyperparameters $P(\bm{\mu}_\psi, \bm{\Sigma}_\psi | \, \mathcal{D}, \mathcal{Z})$ obtained by integrating over Eq. 
\ref{globalposterior}.  This is to be contrasted with simpler approaches that merely find point estimates of the model parameters.   The 
process of prediction uses this posterior probability together with new light curve data $\tilde{\mathcal{D}}_s$ to compute the predictive 
posterior $P( \tilde{\mu}_s | \, \tilde{\mathcal{D}}_s, \mathcal{D}, \mathcal{Z})$.  The marginalization over the hyperparameters correctly 
incorporates the uncertainties in the means, variances, and correlations of the absolute magnitudes and light curve shape parameters.  
Recall that the training set is $\mathcal{D} = \{ \mathcal{D}_s \}$ and $\mathcal{Z} = \{ \mathcal{Z}_s \}$.  The resubstitution distance 
estimate is obtained by setting $\tilde{\mathcal{D}}_s = \mathcal{D}_s$ and computing the predictive probability $P( \mu_s | \, 
\mathcal{D}_s, \mathcal{D}, \mathcal{Z})$ for each supernova $s$ in the training set.

\begin{figure}[t]
\centering
\includegraphics[angle=0,scale=0.45]{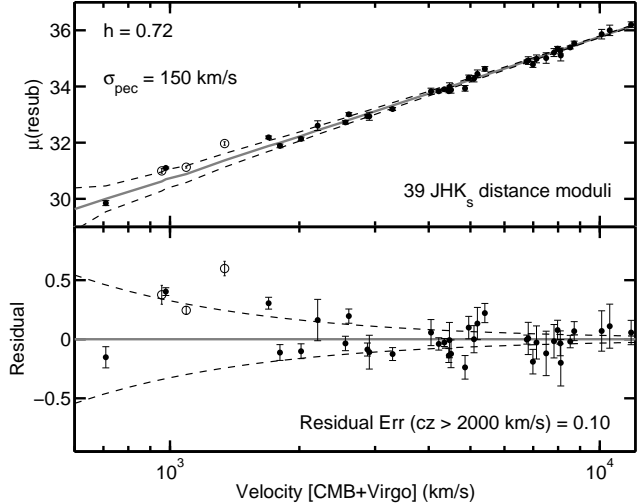}
\caption{\label{hubble_resub} Hubble diagram constructed by resubstitution of training set NIR SN Ia light curves into the trained 
statistical model.   The dotted lines indicate the uncertainty in distance modulus due to peculiar velocities. The average residual at $cz 
> 2000 \text{ km s}^{-1}$ is an excellent 0.10 mag.  The three open circles are the SN with $A_V > 2$ as measured from the optical 
light curves with MLCS2k2.}
\end{figure}

The uncertainty-weighted mean resubstitution error is then a sum over all resubstituted predicted distances for each supernova.
\begin{equation}
\text{err}^2_\text{resub} = \sum_{s=1}^{N_{\text{SN}}} w_s \times \Big[\mu_\text{resub}^s - \mathbb{E}(\mu_s | z_s) \Big]^2 \Big/ 
\sum_{s=1}^{N_{\text{SN}}} w_s
\end{equation}
where the expected predictive distance is $\mu_\text{resub}^s$, the variance is $\sigma^2_{\text{resub},s}$ and the weights are 
$w_s^{-1} = \sigma^2_{\mu,s} + \sigma_{\text{resub},s}^2$.  The resubstitution predictive distance moduli and uncertainties are 
listed in Table 4.   Figure \ref{hubble_resub} shows the Hubble diagram constructed from the resubstitution distance moduli from our 
$JHK_s$ statistical model.   We compute the training resubstitution error over the training set SN with recession velocities $cz > 2000 
\text{ km s}^{-1}$, and find $\text{err}_\text{resub} = 0.10$ mag.

\subsubsection{Cross-Validation and Prediction Error}

The resubstitution error is an optimistic estimate of the predictive, or generalization, error arising from predicting the distances of new 
SN Ia light curves that were not in the training set (``out of sample'').  The resubstitution prediction $P( \mu_s | \, \mathcal{D}_s, 
\mathcal{D}, \mathcal{Z})$ conditions on the data $\mathcal{D}_s$ twice: once during training when it is included in the training set $
\mathcal{D}, \mathcal{Z}$ and once in resubstitution to assess training error.  Double use of the data for both training and evaluation is 
likely to lead to optimistic measures of predictive performance, underestimating the true predictive error.  It is always possible to reduce 
the residuals of a fit to a finite, noisy sample by introducing arbitrarily more complex relations, but arbitrarily complex models will 
typically not generalize well to new data. We should compute the generalization error for out-of-sample cases 
to assess predictive performance of the statistical model.   The distinction between the training error, or Hubble residual, and the 
expected prediction error has not been fully addressed in the literature on SN Ia light curve inference methods.  In this section, we 
describe the novel application of a cross-validation procedure to assess the prediction error and to test sensitivity of the statistical 
model to the training set SN.

To estimate the out-of-sample prediction error and to avoid using the light curve data twice for training and evaluation, we performed 
bootstrap cross-validation \citep{efron83, efron97}.  We sample SN with replacement from the original training set to simulate the 
generation of alternative training sets of the same size.   Because of the random resampling, each bootstrapped training set will 
typically contain duplicate SN data sets and will be missing others.  Each bootstrapped training set of size $n$ SN will be missing 
approximately $(1-1/n)^n \approx 37\%$ of the SN in the original training set.  The SN missing from the bootstrap training set form a 
prediction set, on which we assess the predictive error of a model trained on the complementary training set.  Let $\mathcal{D}^B, 
\mathcal{Z}^B$ be a training set bootstrapped from  the original $\mathcal{D}, \mathcal{Z}$.  Then the prediction set is $\mathcal{D} 
\setminus \mathcal{D}^B$.  To train the statistical model we compute $P(\bm{\mu}_\psi, \bm{\Sigma}_\psi | \, \mathcal{D}^B, 
\mathcal{Z}^B)$ as in Eq. \ref{globalposterior}.  For each supernova light curve $\mathcal{D}_s \in \{\mathcal{D} \setminus 
\mathcal{D}^B\}$, we compute the predictive density $P(\mu_s | \, \mathcal{D}_s, \mathcal{D}^B, \mathcal{Z}^B)$.   This random 
process is repeated  so that each supernova distance is predicted several times from different bootstrapped training sets.  This process 
avoids using each SN light curve simultaneously for both prediction and training.
 
\begin{figure}[t]
\centering
\includegraphics[angle=0,scale=0.45]{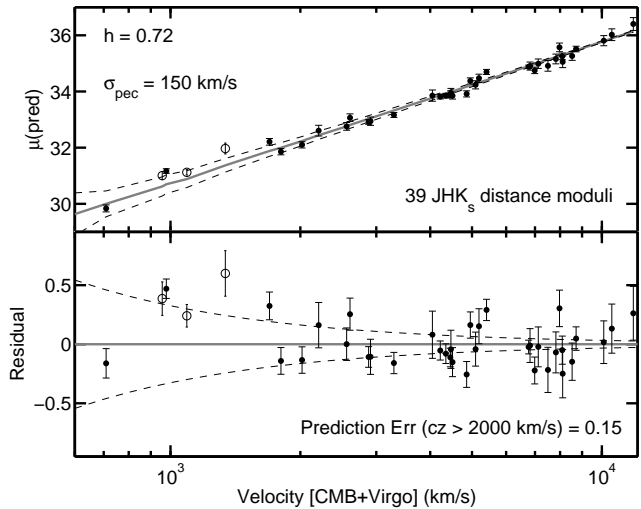}
\caption{\label{hubble_pred} Hubble diagram constructed using predicted distances of NIR SN Ia light curves obtained by inferring the 
statistical model from 50 bootstrapped training sets.  The error bars include both the predictive posterior uncertainty and the scatter 
over multiple bootstrapped predictions.  The estimate of the prediction error for $cz > 2000 \text{ km s}^{-1}$ is an excellent 0.15 mag.  
The three open circles are the SN with $A_V > 2$ as measured from the optical light curves with MLCS2k2.}
\end{figure}

We repeated this process fifty times for the original training set in Table 2.  On average, each supernova is held out of the training set 
and its distance modulus $\mu_s$ is predicted about eighteen times.   For each supernova, the average over all of its predictions $
\bar{\mu}_{\text{pred}}$ and the standard deviation $s_\text{pred}$ over all predictions are listed in Table 4.  We also list the sum of the 
variance over predictions $s^2_\text{pred}$ and the average uncertainty of a prediction (the variance of $P(\mu_s | \, \mathcal{D}_s, 
\mathcal{D}^B, \mathcal{Z}^B)$) as $\sigma_\text{pred}^2$.  Often the uncertainty of a single prediction is larger than the scatter of the 
predictions from different bootstrapped training sets, although this is not always true.
 In Fig. \ref{hubble_pred} we show the Hubble diagram of mean predicted distance moduli $\bar{\mu}_{\text{pred}}$ and their total 
scatter $\sigma_\text{pred}$.  Because we do not use the data twice for training and prediction, the scatter about the Hubble line is less 
tight than in the resubstitution Hubble diagram, Fig. \ref{hubble_resub}.


The ``leave-one-out'' bootstrap error is computed as an uncertainty-weighted average of squared prediction errors.
\begin{equation}
\begin{split}
\text{Err}_{(1)}^2 = \frac{\sum_{B=1}^{50} \sum_{s\in \{\mathcal{D}\setminus\mathcal{D}^B \}} w_s^B \times \Big[ \mu_{\text{pred},B}^s - 
\mathbb{E}(\mu_s | z_s) \Big]^2 }{  \sum_{B=1}^{50} \sum_{s\in \{\mathcal{D}\setminus\mathcal{D}^B \}} w_s^B }
\end{split}
\end{equation}
where  $\mu_{\text{pred},B}^s \equiv \mathbb{E}(\mu_s | \mathcal{D}_s, \mathcal{D}^B, \mathcal{Z}^B)$ and the weights are $
(w_s^B)^{-1} = \sigma_{\mu,s}^2 + \text{Var}[\mu_s | \, \mathcal{D}_s, \mathcal{D}^B, \mathcal{Z}^B]$.  
This bootstrap error estimate is known to be upwardly biased.  \citet{efron83} and \citet{efron97} have shown that a better estimate of 
prediction error is obtained by averaging the bootstrap error with the resubstitution error, using the ``.632 bootstrap estimator'':
\begin{equation}
\text{Err}_{.632}^2 = 0.632 \times \text{Err}_{(1)}^2 + 0.368 \times \text{err}_\text{resub}^2 
\end{equation}
For the Hubble flow SN ($cz > 2000 \text{ km s}^{-1}$) in our sample, we compute this estimate of prediction error: $\text{Err}_{.632} = 
0.15$ mag.  This is a larger error than the resubstitution error computed above, as expected.  However, it is a more realistic estimate of 
predictive performance of distance estimation with our $JHK_s$ light curve model and the current SN sample.  This result confirms that NIR SN Ia are excellent 
standard candles.

This process of resampling of alternative training set tests how sensitive the predictions are to the composition of the finite training set.  
If the statistical model is reasonable, and we had an infinite training set, we would expect the original training set to be representative 
of the
population of NIR SN Ia and we would expect the resampled training sets (and also the complementary held out sets) to look like the
original, and also be representative of the
population.  We would expect that the prediction error and the resubstitution error to be almost the same.  
Our actual training set is finite, so the resampled sets will not look exactly like the original set.  This procedure
tests the sensitivity to the finite sample in addition to making predictions without double use of the light curve data.

The gap between the estimated prediction error (0.15 mag) and the resubstitution error (0.10 mag) tells us that the trained statistical 
model is sensitive to the
finite sampling of the training set.  A larger training set we would be more robust to resampling and we expect that the future predictive 
uncertainty will be in between the current resubstitution error and the estimated predictive error.   A na\"{i}ve argument would suggest 
that if this gap of 0.05 mag between the prediction error and the training error decreases with the square root of the number of SN Ia in 
the sample, then a set of $\sim 200$ SN Ia would reduce it to about 0.02 mag, and a few hundred would be needed to reduce it to $
\sim$ 0.01 mag.   To build up statistical strength and further solidify our knowledge of the properties of SN Ia in the NIR,  we are 
continuing our campaign to observe SN Ia in the near infrared with PAIRITEL.

\subsection{The Effect of Dust on the SN Ia NIR Sample}\label{dustsection}

The results presented thus far have ignored the effects of dust extinction in the NIR sample.   We can examine the possibility of 
extracting information about the dust distribution from the NIR by looking at the colors $J_0-H_0$, $J_0-K_{s0}$, and $H_0-K_{s0}$ at 
peak.  From the population hyperparameters we can compute the mean and standard deviation of these colors.  The mean colors are 
-0.25, 0.0 and 0.25, and their population dispersions are     0.14, 0.20, 0.17, respectively.  This means that, for optical extinctions that 
are less than $A_V \sim 2-3$, dust extinction in the near infrared cannot be clearly distinguished from intrinsic color variations, 
because of the diminished effects of dust absorption in the NIR.  Only SN with much greater NIR reddening carry information on the 
dust distribution from their NIR data alone.  Estimates of the optical $A_V$ extinctions of the SN in our sample from MLCS analysis of 
optical light curves \citep{hicken09a} were reported in WV08.  There are only three highly reddened SN with $A_V > 2$: SN 1999cl, 
2006X, and SN 2003cg ($A_V = 3.49, 3.83, 4.20$, respectively) in the sample of 39 SN, and they are depicted in Figs. 
\ref{hubble_resub},  \ref{hubble_pred} with open circles.  Although they are redder in their NIR colors than the population mean, their 
colors are only about $\sim1-2\sigma$ redder, so they are barely distinguishable from the intrinsic color variations.  This conclusion 
does not change if we calculate the mean and standard deviation of the colors by including or excluding the highly optically reddened 
SN.

If we take the $A_V$ estimates from the optical data at face value, we can estimate the likely effect of dust extinction on our posterior 
estimates of absolute magnitudes.  The relative weight of a particular supernova in posterior inferences about absolute magnitude-
related quantities (means, variances and correlations) is inversely proportional to its magnitude uncertainty due to peculiar velocities: 
$w_M^s = c/\sigma_{\mu,s}^2$ and $c^{-1} =  \sum_s \sigma^{-2}_{\mu,s}$.  Comparing these weights to the $A_V$ for each SN, we 
find that 87\% of the magnitude weight lies with SN with $A_V < 0.5$, 97\% of the weight is in $A_V < 1$, and 99.7\% of the weight is in 
$A_V < 2$.  The three SN with $A_V > 2$ have a total weight of 0.32\% in magnitude calculations.  Although the highly 
reddened supernovae have large Hubble residuals in Figs. \ref{hubble_resub},  \ref{hubble_pred}, since they are at low redshifts 
where the contribution of peculiar velocity uncertainties to their distance uncertainty is large, they have little influence on the posterior 
inferences about the NIR absolute magnitudes.  Furthermore,  they have no effect on the estimates of the training error or prediction 
error, because only the Hubble flow SN at $cz > 2000 \text{ km s}^{-1}$ are useful for validation of the statistical model.

The weighted mean $A_V$ value of the sample is $\sum_s w_M^s A_V^s = 0.23$ mag.  Assuming a CCM law with $R_V = 2$, this 
means that the estimated NIR absolute magnitudes would be impacted by mean extinctions of about $A_J = 0.05$, $A_H = 0.03$ and 
$A_{K} = 0.02$.  The weighted scatters in the NIR extinctions implied by the $A_V$ values are about $\sigma(A_J) = 0.08$, $
\sigma(A_H) = 0.05$, $\sigma(A_K) = 0.03$.  If these are the dust contributions to the measured dispersions $\sigma(M_X)$, then 
subtracting them in quadrature yields intrinsic dispersions of $\sigma(M_J) = 0.15$, $\sigma(M_H) = 0.10$, and $\sigma(M_{Ks}) = 
0.19$.  These rough estimates do not substantially change our results.  

We conclude that the NIR sample alone contains little if any information about the dust distribution and hence it is not worthwhile to 
use the full model described in \S 2 at this time to infer the dust properties.   Additionally, if we extrapolate the $A_V$ estimates 
obtained from the optical data to the near infrared, the estimated effect of dust extinction is fairly small.  These rough estimates do not 
take into account the non-gaussianity of the dust distribution and a full Bayesian analysis of the directed graph in Fig. \ref{dag}, 
conditioned on both the NIR and optical data simultaneously will be required to obtain informative inferences about the dust 
properties (Mandel et al. 2009, in prep.).

\section{Conclusion}
 
We have constructed the hierarchical Bayesian formulation of statistical inference with SN Ia light curves, and represented the 
probabilistic structure using formal graphical models.  Furthermore, we have presented a Markov Chain Monte Carlo algorithm that 
uses the conditional independence structure of the equivalent directed acyclic graph to efficiently sample the global posterior 
probability distribution over individual light curve parameters and population hyperparameters for  training the statistical model on the 
low-$z$ data set, and for prediction on future SN Ia data.   We have applied this approach and computational method to the $JHK_s$ 
light curve data set compiled by WV08, including a recent homogeneous set of light curves from PAIRITEL, and computed the joint 
posterior probabilities over all individual light curve parameters (Table 2) and the statistical characteristics of the population, including 
the covariance of absolute magnitudes and $J$-band light curve shape parameters (Table 3).
 
We summarize the assumptions of our statistical model.  First, we have assumed that the normalized $H$- and $K_s$- band light 
curves of different SN are identical between -10 and 20 days around maximum.  Furthermore, we have posited a parametric light curve 
model for the $J$-band between -10 and 60 days that captures the variations in the double-peaked structure.  A quick look at the data 
and the template models we have constructed in Fig. \ref{HKtemp} and Fig. \ref{Jcomparison} reveals that these are  reasonable 
models for the $JHK_s$ data.  The major assumption in the application of our hierarchical model is that the parameters 
governing the multi-band absolute light curves are drawn from a jointly multivariate Gaussian population distribution.   This is the 
simplest multivariate distribution that models correlations, and its use is reasonable in the absence of other guiding information.  Our 
results (Figs. \ref{mvJHK}-\ref{MKvsRise}) reveal no obvious deviations from this assumption, but this is certainly not proof, especially 
with a small sample.  This assumption must be constantly re-evaluated in applications of the hierarchical framework to larger or 
different data sets or with other light curve models.  In this paper, we have not estimated the dust-related aspects of Fig. \ref{dag}, 
because the effects of dust are small for our NIR sample.  However, in future studies in conjunction with optical data, the full graph can 
be computed using \textsc{BayeSN} to perform probabilistic inference of the SN Ia population and dust distribution.

The marginal intrinsic scatter in peak absolute magnitudes were found to be  $\sigma(M_J) = 0.17 \pm 0.03$, $\sigma(M_H) = 0.11 \pm 
0.03$, and $\sigma(M_{Ks}) = 0.19 \pm 0.04$. 
We have presented the first quantitative measurements of the correlations of NIR absolute magnitudes with $J$-band light curve 
shape.   We showed that with greater than 95\% probability there are positive correlations between peak $JHK_s$ absolute 
magnitudes and the $J$-band post-trough rise rate.  Intrinsically dimmer SN Ia light curves tend to rise to the second $J$-band 
maximum faster.  Since in our $J$-band model, the post-second-peak decline rate is linked to rise rate, this also suggests that the late-
time slopes of $J$-band light curves are steeper for dimmer SN.   We have also quantitatively measured correlations of the rise rate 
with other aspects of the light curve shape (Table 3), which show that faster decline rates go with faster rise rates, shorter times to 
trough and shorter times to the second maximum.
These results suggest that  NIR SN Ia are excellent standard candles at peak, and they can  be improved by using the information in the 
late-time light curve.  

These relations may be useful for better understanding of SN Ia progenitor explosions in conjunction with physical modeling.  The 
theoretical models of \citet{kasen06} suggest that the structure of the secondary maximum in the NIR is related to the ionization 
evolution of the iron group elements in the SN atmosphere.  They also indicate that NIR peak absolute magnitudes have relatively 
weak sensitivity to the input progenitor $\,^{56}$Ni mass, with a dispersion of $\sim 0.2$ mag in $J$ and $K$, and $\sim 0.1$ mag in 
$H$ over models ranging from 0.4 to 0.9 solar masses of $\,^{56}$Ni.   The optical and bolometric peak magnitudes have much larger 
variations over the same range of mass.  Further observational studies of SN Ia in the NIR may place valuable constraints on 
theoretical explosion models.
 
We constructed a Hubble diagram with the training set SN Ia, and found an average residual of 0.10 mag for $cz > 2000 \text{ km 
s}^{-1}$.  We have also performed bootstrap cross-validation to estimate the out-of-sample prediction error, which was found to be an 
excellent 0.15 mag.  The gap between these estimates suggests that a larger sample  is needed to solidify our 
inferences about the population of near-infrared SN Ia light curves.  Our group continues to collect an extensive set of nearby NIR SN 
Ia light curves using PAIRITEL.  With an ever-growing data set, in the near future,  we will be able to expand the model considered 
here to include more extensive light curve models in $H$ and $K_s$, and to combine the NIR and optical data to gain a better 
understanding of SN colors and dust extinction (Mandel et al. 2009, in prep.).

It is worth considering whether the propitious properties of SN Ia in the NIR can be leveraged by future space missions for SN Ia 
cosmology.  The measurement of dark energy properties by the NASA/DOE Joint Dark Energy Mission will be limited by systematic 
effects, in particular dust extinction.  The diminished absorption by dust and the narrow dispersion of peak luminosities in the NIR, 
particularly in the $H$-band, may be crucial to the precise measurement of dark energy, if observations of high-$z$ SN can be 
conducted in the rest-frame NIR.

 \acknowledgements
 
K.M. thanks St\'{e}phane Blondin, Peter Challis, Jonathan Chang, Ryan Foley, Andrew Gelman, Malcolm Hicken, Joseph Koo, Sam 
Kou, and Gautham Narayan for useful discussions and clarifications.  We thank the anonymous referee for useful 
suggestions that led to an improved manuscript.  Supernova research at Harvard University is supported by NSF grant  AST06-06772.  
The Peters Automated Infrared Imaging Telescope (PAIRITEL) is operated
by the Smithsonian Astrophysical Observatory (SAO) and was made
possible by a grant from the Harvard University Milton Fund, the
camera loan from the University of Virginia, and the continued support
of the SAO and UC Berkeley. Partial support for PAIRITEL operations 
comes from National Aernonautics and Space Administration
(NASA) grant NNG06GH50G (``PAIRITEL: Infrared Follow-up for Swift  
Transients''). This publication makes use of data products from the 2MASS Survey, funded
by NASA and the US
National Science Foundation (NSF). IAUC/CBET were useful. M.W.V. is funded
by a grant
from the US National Science Foundation (AST-057475). A.S.F. acknowledges
support from
an NSF Graduate Research Fellowship and a NASA Graduate Research Program
Fellowship.

\appendix 

\section{A. Conditional Independence and D-Separation}\label{dsep}

Let $P( \{ \theta_i \})$ be a joint distribution of random variables represented by a directed acyclic graph.  Consider three disjoint 
subsets of the random variables $\{ \theta_i \}:  \mathcal{A}, \mathcal{B}$ and $\mathcal{C}$.  Two sets are \emph{marginally 
independent} if $P(\mathcal{A}, \mathcal{B} ) = P(\mathcal{A}) P(\mathcal{B})$.  Two sets, $\mathcal{A}$ and $\mathcal{B}$, are 
\emph{conditionally independent} given a third set $\mathcal{C}$ if $P(\mathcal{A} , \mathcal{B}| \mathcal{C}) = P(\mathcal{A} | 
\mathcal{C} ) P(\mathcal{B} | \mathcal{C})$.  Marginal independence between $\mathcal{A}$ and $\mathcal{B}$ can be seen in a 
directed graph because there will be no links between the nodes in set $\mathcal{A}$ and the nodes in set $\mathcal{B}$.   
Conditional independence indicates that if the values of the nodes in $\mathcal{C}$ are known, then the variables in $\mathcal{A}$ 
and those in $\mathcal{B}$ are statistically independent.   Graphically, this means that all directed or undirected paths (ignoring the 
arrows) from one set to the other are ``blocked'' by nodes in $\mathcal{C}$.  

Conditional independence between two sets of nodes 
given a third set can be ascertained from a directed graph using the d-separation property \citep{pearl88}:  A path between a node in $\mathcal{A}$ and a node in $\mathcal{B}$ is blocked at node $\theta_i$ if (1)  the intermediate node $\theta_i$ is in set $\mathcal{C}$ and the arrows meet at $\theta_i$ in a tail-to-tail or head-to-tail fashion (not convergent), or (2) the arrows meet head-to-head (convergent) and the intermediate node $\theta_i$ is not in $\mathcal{C}$, and neither are any of its descendants.  
The nodes $\mathcal{A}$ are d-separated from the nodes $\mathcal{B}$ given set $\mathcal{C}$ if all paths between elements in $
\mathcal{A}$ and $\mathcal{B}$ are blocked.  If the nodes $\mathcal{A}$ are d-separated from the nodes $\mathcal{B}$ by $
\mathcal{C}$, then $\mathcal{A}$ is conditionally independent from $\mathcal{B}$ given $\mathcal{C}$.

\section{B. The \textsc{BayeSN} Algorithm - Mathematical Appendix}\label{bayesnmath}

In this appendix, we present mathematical details of the \textsc{BayeSN} algorithm.  Let $\bm{\psi}_s^{-M_0^F}$, $\bm{\psi}_s^{-\text{L},F}$, 
and $\bm{\psi}_s^{-\text{NL},F}$ indicate all the intrinsic parameters in $\bm{\psi_s}$ other than the peak absolute magnitude, the 
linear shape parameters, and the nonlinear shape parameters in filter $F$, respectively.

1.   We have used the conjugate hyperprior density $P(\bm{\mu}_\psi, \bm{\Sigma}_\psi )$ defined in Eqns. \ref{hyperprior} and 
\ref{hyperprior2} and choose the noninformative limit by setting $\kappa_0 = 0, \nu_0 = -1$, and $\bm{\Lambda}_0 = \epsilon_0 \bm{I}$ 
for small $\epsilon_0$.   Let $\bar{\bm{\psi}}$ be the sample mean of the $\{ \bm{\psi}_s\}$, and let 
$\bm{S}_\psi = \sum_{s=1}^{N_{\text{SN}}} (\bm{\psi}_s - \bar{\bm{\psi}}) (\bm{\psi}_s - \bar{\bm{\psi}})^T$
be the matrix sum of squared deviations from the mean.  The conditional posterior density $P(\bm{\mu}_\psi, \bm{\Sigma}_\psi | \, \{ \bm{\psi}_s\})$ can be decomposed as 
$\bm{\Sigma}_\psi | \, \{ \bm{\psi}_s\} \sim \text{Inv-Wishart}_{N-1}\left( [\bm{\Lambda_0} + \bm{S}_\psi]^{-1}\right)$ and
$\bm{\mu}_\psi | \, \bm{\Sigma}_\psi , \{ \bm{\psi}_s\} \sim N( \bar{\bm{\psi}},  \bm{\Sigma}_\psi/N_{\text{SN}})$
\citep{gelman_bda}.  We first directly sample a new covariance matrix $\bm{\Sigma}_\psi$ from the inverse Wishart distribution  The 
matrix drawn in this way is guaranteed to be a proper covariance matrix (i.e. positive semi-definite).  Conditional on that matrix we 
directly sample a new population mean $\bm{\mu}_\psi$ from the multivariate normal distribution.

2.  Let $\bar{A}$ be the sample mean of the $\{ A_H^s \}$.  The conditional posterior density is $P(\tau_A | \, \{A^s_H\} ) = \text{Inv-
Gamma}(N_{\text{SN}}-1, N_{\text{SN}} \bar{A})$.

3a.   Let $N^{-1} = \bm{1}^T  \bm{W}^{-1}  \bm{1}$ and
$\bar{F}_0 = N \bm{1}^T (\bm{W}_s^F)^{-1}  [\bm{m}^F_s - \bm{L}_0^F(\bm{\theta}^F_{\text{NL},s}) -\bm{L}_1^F(\bm{\theta}^F_{\text{NL},s}) 
\bm{\theta}_{\text{L},s}^F ]$. 
We can compute the population conditional expectation: $\tilde{F}_0 = \mu_s + A^F_s + \mathbb{E}[ M_{0,s}^F | \bm{\psi}_s^{-M_0^F}, 
\bm{\mu}_\psi, \bm{\Sigma}_\psi ]$ and the population conditional variance $C  = \text{Var}[M_{0,s}^F |  \, \bm{\psi}_s^{-M_0^F}, 
\bm{\mu}_\psi, \bm{\Sigma}_\psi ]$, using the conditioning property of the multivariate Gaussian distribution.  Then the conditional 
density of $F_{0,s}$ is normal
$N( F_{0,s} | \, \hat{F}_0, \Lambda)$ with variance $\Lambda = (N^{-1} + C^{-1})^{-1}$ and mean $\hat{F}_0 = \Lambda (N^{-1} 
\bar{F}_0 + C^{-1} \tilde{F}_0)$.  

3b.  Compute $\bm{N}^{-1} =  \bm{L}_1^{F,T}(\bm{\theta}^F_{\text{NL},s}) (\bm{W}_s^F)^{-1} \bm{L}_1^F(\bm{\theta}^F_{\text{NL},s})$ and
$\bar{\bm{\theta}}_{\text{L}}^F = \bm{N} \bm{L}_1^{F,T}(\bm{\theta}^F_{\text{NL},s}) (\bm{W}_s^F)^{-1} [ \bm{m}^F_s - \bm{1} F_{0,s} - 
\bm{L}_0^F(\bm{\theta}^F_{\text{NL},s})]$.
The conditional population expectation and variance are: $\tilde{\bm{\theta}}^F_{\text{L}} = \mathbb{E}[ \bm{\theta}_{\text{L},s}^F | \, 
\bm{\psi}_s^{-\text{L},F}, \bm{\mu}_\psi, \bm{\Sigma}_\psi ]$ and 
$\bm{C}  = \text{Var}[  \bm{\theta}_{\text{L},s}^F | \, \bm{\psi}_s^{-\text{L},F}, \bm{\mu}_\psi, \bm{\Sigma}_\psi ]$.
The conditional posterior density of $\bm{\theta}^F_{\text{L},s}$ is normal $N( \bm{\theta}^F_{\text{L},s} | \, \hat{\bm{\theta}}_{\text{L}}^F, 
\bm{\Lambda})$
with covariance matrix $\bm{\Lambda} = (\bm{N}^{-1} + \bm{C}^{-1})^{-1}$ and mean $\hat{\bm{\theta}}_{\text{L}}^F = \bm{\Lambda} 
( \bm{N}^{-1} \bar{\bm{\theta}}^F_{\text{L}} + \bm{C}^{-1} \tilde{\bm{\theta}}^F_{\text{L}})$.  Note that steps 3a and 3b could be combined 
by Gibbs sampling from $P(F_{0,s}, \bm{\theta}^F_{\text{L},s} | \, \bm{\phi}_s^{-\text{L},F}, \mu_s, \bm{A}_s; \bm{\mu}_\psi, \bm{\Sigma}_
\psi, \tau_A, \mathcal{D}_s, z_s)$.

3c.  Compute the expectation and covariance of the conditional population density:
$\tilde{\bm{\theta}}^F_{\text{NL}} = \mathbb{E}[ \bm{\theta}_{\text{NL},s}^F | \, \bm{\psi}_s^{-\text{NL},F}, \bm{\mu}_\psi, \bm{\Sigma}_
\psi ]$ and $\bm{C}  = \text{Var}[ \bm{\theta}_{\text{NL}}^F  | \, \bm{\psi}_s^{-\text{NL},F}, \bm{\mu}_\psi, \bm{\Sigma}_\psi ]$.
The conditional posterior density of the nonlinear parameters in band $F$,  $\bm{\theta}_{\text{NL}}^F$ is
proportional to $ N( \bm{m}_s^F | \,  \bm{1}F_{0,s} + \bm{L}_0^F( \bm{\theta}_{\text{NL},s}^F) + \bm{L}_1^F( \bm{\theta}_{\text{NL},s}^F) 
\bm{\theta}_{\text{L}}^F, \bm{W}_s^F)
\times N(  \bm{\theta}_{\text{NL},s}^F | \, \tilde{\bm{\theta}}_{\text{NL}}^F, \bm{C}).$
We obtain a proposal $\bm{\theta}_{\text{NL},s}^{F,*} \sim N( \bm{\theta}_{\text{NL},s}^F,  \bm{\Sigma}_{\text{jump},s}^{\text{NL},F} )$, 
and apply the Metropolis rejection rule.

3d.  We allow for the possibility that the probability density of the distance modulus conditioned on the redshift only, $P(\mu_s | z_s)$, 
may be mildly non-Gaussian.  The conditional posterior density is $P(\mu_s | \bm{\phi}_s, \bm{A}_s; \bm{\mu}_\psi, \bm{\Sigma}_\psi , 
\mathcal{D}_s, z_s) \propto N(\bm{\phi}_s - \bm{A}_s - \bm{v}\mu_s | \, \bm{\mu}_\psi, \bm{\Sigma}_\psi) \times P(\mu_s | z_s)$, and 
generally cannot be sampled directly.  However, we can approximate $P(\mu_s | z_s) \approx N(\mu_s | \mu_g \equiv f(z_s), 
\sigma^2_\mu)$ with a Gaussian using Eq. \ref{gaussianmuprior}.    We choose the Metropolis-Hasting proposal distribution to be 
$Q(\mu_s^* | \bm{\phi}^s, \bm{A}_s, \bm{\mu}_\psi, \bm{\Sigma}_\psi) \propto N(\bm{\phi}_s - \bm{A}_s - \bm{v}\mu_s^* | \, \bm{\mu}_
\psi, \bm{\Sigma}_\psi) \times N(\mu_s^* | \mu_g, \sigma_{\mu}^2) = N( \mu_s^* | \, \hat{\mu}, \hat{\sigma}^2_\mu)$, where $
\hat{\sigma}^2_\mu = (\sigma_{\mu}^{-2} + s^{-2})^{-1}$,  and $\hat{\mu} = \hat{\sigma}_\mu^{-2} ( \sigma_{\mu}^{-2} \mu_g + s^{-2} 
\tilde{\mu})$ is a weighted average of the distance information from the redshift and the light curves.  The mean $\tilde{\mu} = s^2  
\bm{v}^T \bm{\Sigma}_\psi^{-1} (\bm{\phi}_s - \bm{\mu}_\psi - \bm{A}_s)$ and variance $s^2 = (\bm{v}^T \bm{\Sigma}_\psi^{-1} 
\bm{v})^{-1}$ describe the distance information from the individual SN light curves only.  We draw a proposed $\mu_s^*$ from $Q$.
The Metropolis-Hastings ratio is computed from Eq. \ref{mhratio} using the above conditional posterior density and the proposal density.  After cancellation of terms, this simplifies to: $r = P(\mu_s^* | z_s) N(\mu_s  | \, \mu_g, \sigma^2_{\mu}) / P(\mu_s | z_s) 
N(\mu^*_s | \, \mu_g, \sigma^2_{\mu})$.
If  $P(\mu_s | z_s)$ is actually close to Gaussian, Eq. \ref{gaussianmuprior}, then the M-H ratio is identically one, and the proposal $
\mu^*_s$ is  always accepted, as this is the same as Gibbs sampling.  If $P(\mu_s | z_s)$  is mildly non-Gaussian, then $r$ and the acceptance 
rate will be slightly less than one.

3dP.  For distance prediction, the distance modulus $\mu_s$ is Gibbs sampled from the conditional posterior density $N(\mu_s | \, 
\tilde{\mu}, s^2)$. 

3e. Sample a proposed extinction $A_H^{s,*} \sim N(A_H^s, \sigma^2_{\text{jump,A,s}})$.  The conditional posterior density of the 
extinction $A_H^{s,*}$ is proportional to $N( \bm{\phi}_s - \bm{v} \mu_s - \bm{A}_s(A_H^{s,*}) | \bm{\mu}_\psi, \bm{\Sigma}_\psi) \times 
\text{Expon}( A_H^{s,*} | \, \tau_A)$.  Apply Metropolis rejection.


\section{C. \textsc{BayeSN} - Practical Considerations}\label{practical}

The chain is seeded with initial starting estimates for all the parameters.  It is useful before running the MCMC to obtain rough point 
estimates of the light curve parameters using, e.g. the maximum likelihood estimate (MLE).  Point estimates of the distance moduli can 
be obtained from $\mathbb{E}(\mu_s | \, z_s)$.  The extinction values $A_H^s$  can be chosen to be small random numbers.   Random 
noise is added to these point estimates to generate different starting positions of each chain, to ensure that each chain begins in a 
different region.

The Metropolis steps within the Gibbs scan use jumping kernels that must be  tuned to generate efficient MCMC chains.  The scalar 
kernels $\sigma_{\text{jump},A,s}$ are tuned to generate $\sim 40\%$ acceptance rates for their respective Metropolis steps.  This is 
easily done by running a few preliminary short chains to compute the average acceptance rates and adjusting the jumping sizes 
accordingly.  The nonlinear jumping kernel, $\bm{\Sigma}_{\text{jump},s}^{\text{NL},F} $ is a matrix if there are more than one 
nonlinear parameters in the light curve model for band $F$.   This can be estimated from the inverse of the Fisher information matrix at 
the MLE estimate, or from the sample covariance of the $\bm{\theta}_{\text{NL},s}^F$ chain values, to reflect the shape of the 
underlying density. The overall size of the matrix is then scaled to produce acceptance rates of $\sim 40\%$ in preliminary short runs, 
or $\sim 23\%$ if the dimensionality of $\bm{\theta}_{\text{NL},s}^F$ is high \citep{gelman_bda}.   Once the jumping kernels have been 
set to appropriate values, long chains are run.

To assess the convergence of the MCMC, a few independent long chains with different initial positions are run.  The 
\text{BayeSN} computation is easily be parallelized as each independent MCMC chain can be run on a separate processor.  
The Gelman-Rubin statistic \citep{gelman92} compares the between-chain variances with the within-chain variances in each 
parameter  to compare the coverages of the chains .  If the chains have converged, the Gelman-Rubin ratio should be close to 1.  The 
sample paths of representative parameters are inspected visually to ascertain that the chains are well mixed.   Upon convergence, the 
initial portions of each chain are discarded as ``burn-in'', and the chains are concatenated for final inferences.

\bibliographystyle{apj}
\bibliography{apj-jour,sn}{}

\end{document}